\documentclass[amsmath,aps,amssymb,showkeys,superscriptaddress,nofootinbib,10pt]{revtex4}

\usepackage{graphicx,color}
\usepackage{subfigure}
\usepackage{setspace}

\usepackage{amsmath,amssymb,ascmac}
\usepackage{type1cm}
\usepackage{bm}

\begin{document}

%=================================================================

\title{Chiral Symmetry in Dense Matter with Meson Condensation}

\author{Takumi Muto}
\email{takumi.muto@it-chiba.ac.jp}
\affiliation{Department of Physics, Chiba Institute of Technology, 2-1-1 Shibazono, Narashino, Chiba 275-0023, Japan }

\author{Toshiki Maruyama}
\email{maruyama.toshiki@jaea.go.jp }
\affiliation{Advanced Science Research Center, Japan Atomic Energy Agency , Ibaraki 319-1195, Japan}

\author{Toshitaka Tatsumi}
\email{tatsumitoshitaka@gmail.com}
\affiliation{52-4 Kitashirakawa Kamiikeda-Cho,Kyoto 606-8287,Japan}

\date{\today}

\begin{abstract}
Kaon condensation in hyperon-mixed matter [($Y$+$K$) phase], 
which may be realized in neutron stars, is discussed on the basis of chiral symmetry. 
With the use of the effective chiral Lagrangian for kaon--baryon and kaon--kaon interactions; 
coupled with the relativistic mean field theory and universal three-baryon repulsive interaction, 
we clarify the effects of the $s$-wave kaon--baryon scalar interaction simulated by the kaon--baryon sigma terms 
and vector interaction (Tomozawa--Weinberg term) on kaon properties 
in hyperon-mixed matter, the onset density of kaon condensation, and the equation of state with the ($Y$+$K$) phase.
In particular, the quark condensates in the ($Y$+$K$) phase are obtained, 
and their relevance to chiral symmetry restoration is discussed.
\end{abstract}


\maketitle

\section{Introduction}
The possible existence of meson condensation (MC) in dense and hot hadronic matter has been extensively considered from the viewpoints of particle physics, nuclear physics, astrophysics, and condensed matter physics. 
Much attention has been paid mainly to pion condensation~\cite{sawyer1972,scalapino1972,migdal1978,migdal1990,baym1979,kmttt1993} and kaon condensation~\cite{kn86,t88,mt92,m93,mtt93,mttt1993,tpl94,kvk95,lbm95,lee1996,prakash1997,tstw98,fmmt1996,g2001}.
Meson condensation is characterized as the macroscopic realization of the Nambu--Goldstone (NG) bosons  (specifically pions and kaons) in a strongly interacting system of bosons and baryonic matter. 
Therefore, the meson--baryon dynamics associated with underlying chiral symmetry and its spontaneous or  explicit symmetry breaking play an important role {in} revealing the properties of the meson-condensed state. 
In this respect, the meson-condensed system offers a remarkable contrast to usual condensed matter systems like dilute Bose gases, where  formation of the Bose--Einstein condensation (BEC) occurs as a result of a competition between thermal fluctuation and quantum statistics. 

{In relation to the thermal evolution of neutron stars, it has been suggested that extraordinary rapid cooling processes through neutrino emissions may become possible in the presence of MC~[for example,~refs.\cite{maxwell1977,mtt93} for pion condensation (PC) and refs. \cite{t88,bkpp1988,fmtt1994,fmtt1994-2} for kaon condensation (KC)]. The relevant weak reactions can be described in a unified way together with meson (M)--baryon (B) dynamics on the basis of chiral symmetry.}

Along with the development of strangeness nuclear physics involving kaons, $\phi$ mesons, and hyperons, a possible existence of {KC} has been suggested as a novel hadronic phase with multi-strangeness~\cite{kn86}. 
 It has been shown that the $s$-wave KC is formulated model-independently with a framework of current algebra and {the partial conservation of axial-vector current} (PCAC)~\cite{t88} {in the context of both the EOS and weak reactions}.
Subsequently, a coexistent phase of KC and hyperon ($Y$)-mixed matter [($Y$+$K$) 
phase] has been considered in the relativistic mean field (RMF) theory~\cite{ekp95,kpe95,sm96}, in the effective chiral Lagrangian for kaon ($K$)--baryon ($B$) interaction~\cite{pal2000,m2008,mishra2010}, density-dependent RMF theory~\cite{cb2014,mbb2021,ma2023}, and an effective chiral Lagrangian coupled to the RMF and three-baryon repulsion~\cite{mmt2021,mmt2022,muto2024}.  
The driving force of kaon condensation is brought about by both the $s$-wave $K$-nucleon ($N$) scalar interaction simulated by the $KN$ sigma term $\Sigma_{KN}$ and the $s$-wave $K$--$N$ vector interaction corresponding to the Tomozawa--Weinberg term.  $\Sigma_{KN}$ not only specifies the scale of explicit breaking of chiral symmetry, but also is related with the $\bar q q$ quark {contents} in the nucleon. The onset density of KC, $\rho_{\rm B}(K^-)$, has been estimated to be $\rho_{\rm B}(K^-) $ = (3--4)~$\rho_0$ 
with the nuclear saturation density $\rho_0$ (=0.16~fm$^{-3}$), depending on the value of $\Sigma_{KN}$. 
Beyond the onset density, the kaon-condensed phase in hadronic matter develops, accompanying the softening of the equation of state (EOS) at high densities, and it is eventually considered to move to a chiral-restored phase. {Thus,} the KC may be regarded as a pathway from hadronic matter to strange quark matter and may affect properties of the $\bar q q$ condensate in dense matter, which is an order parameter of chiral restoration. 

In this paper, we clarify the roles of the quark ($\bar q q$) contents in the baryon and quark condensates in the ($Y$+$K$) phase comprehensively in the context of chiral symmetry and its spontaneous and explicit breaking.
First, we reanalyze the allowable range of the {$KB$} sigma term for baryon {$B$} in terms of recent constraints of the $\pi$-$N$ sigma term, $\Sigma_{\pi N}$, and the strangeness contents inside the nucleon, $\langle \bar s s\rangle_N$, which have been obtained from phenomenological analyses and lattice QCD.
Second, the $\bar q q$ condensate in the (Y+K) phase is obtained with the help of the Feynman--Hellmann theorem in the mean field approximation. We discuss the relevance of the $s$-wave KC to chirally restored quark matter through the behavior of the quark condensate as a mediating order parameter between the ($Y$+$K$) phase and quark phase. 

As a background of our present study, we overview our interaction model for the ($Y$+$K$) phase, which has been explored in a series of our works, and summarize the results on the onset density of KC in hyperon-mixed matter, the EOS, and the characteristic features of the ($Y$+$K$) phase~\cite{muto2024}.

The paper is organized as follows. In Section~\ref{sec:kaon}, the chiral symmetry approach for kaon condensation based on the effective chiral Lagrangian {(abbreviated to ChL)} is introduced. 
In Section~\ref{sec:Bforce}, the {``minimal'' RMF (MRMF)} theory is explained in the meson-exchange picture for $B$--$B$ interaction. In addition, the {universal three-baryon repulsion (UTBR) [string junction model (SJM) as a specific model]} and {three-nucleon attraction (TNA)} are introduced phenomenologically. 
The formulation obtaining the ground state energy for the ($Y$+$K$) phase is described in Section~\ref{sec:ground-state}. 
In Section~\ref{sec:SNM}, the results on the properties of the symmetric nuclear matter (SNM) with our interaction model are  given. 
 In Section~\ref{sec:qcb}, the ``$K$$B$ sigma terms'' are estimated by the inclusion of the nonlinear effect with respect to the strange quark mass beyond chiral perturbation in the next-to-leading order.  
In Section~\ref{sec:KP}, the onset of the $s$-wave KC in hyperon-mixed matter and the composition of matter in the ($Y$+$K$) are figured out. 
 In Section~\ref{sec:MR}, the static properties of neutron stars with the ($Y$+$K$) phase such as gravitational mass and radius are summarized.  
Quark condensates in the ($Y$+$K$) phase and relevance to chiral restoration are discussed 
in Section~\ref{sec:qcondensates}. 
A summary and outlook are given in Section~\ref{sec:summary}.

\section{Chiral Symmetry Approach for Kaon Condensation}
\label{sec:kaon}  

The ($Y$+$K$) phase is composed of kaon condensates and hyperon-mixed baryonic matter together with  leptons, being kept in beta equilibrium, charge neutrality, and baryon number conservation. 
{Among the $\Sigma^-$, $\Sigma^0$, and $\Sigma^+$ hyperons and $\Xi^-$, $\Xi^0$ hyperons, $\Sigma^-$ and $\Xi^-$ hyperons are considered as favorable to appear in matter, since the degenerate energy of negatively charged leptons ($e^-$, $\mu^-$) helps meet the onset condition of leptonic weak reactions, although there is another ambiguity concerning the hyperon potential in matter.} In the following, we simply take into account protons, neutrons, $\Lambda$, $\Sigma^-$, and $\Xi^-$ hyperons for baryons and electrons and muons for leptons. 

\subsection*{Kaon--Baryon and Multi-Kaon Interactions}
\label{subsec:kb}

We base our model for $K$--$B$ and $K$--$K$ interactions upon the effective chiral SU(3)$_{\rm L}$ $\times$ SU(3)$_{\rm R}$ Lagrangian~\cite{kn86} in the next-to-leading order $O(p^2)$ with the typical energy scale $p$ in chiral perturbation. The relevant Lagrangian density is given by the following:

\vspace{-12pt}
\begin{eqnarray}
{\cal L}_{K,B}&=&\frac{1}{4}f^2 \ {\rm Tr}( 
\partial^\mu U^\dagger\partial_\mu U) 
+\frac{1}{2}f^2\Lambda_{\chi {\rm SB}}({\rm Tr}M(U-1)+{\rm h.c.}) \cr
&+&{\rm Tr}\overline{\Psi}(i{\gamma^\mu\partial_\mu}-M_{\rm B})\Psi 
+{\rm Tr}\overline{\Psi} \gamma^\mu\lbrack V_\mu, \Psi\rbrack 
+ D {\rm Tr}\overline{\Psi}\gamma^\mu \gamma^5\lbrace A_\mu, \Psi\rbrace
+F {\rm Tr}\overline{\Psi}\gamma^\mu \gamma^5\lbrack A_\mu, \Psi\rbrack \cr
&+& a_1{\rm Tr}\overline{\Psi}(\xi M^\dagger\xi+{\rm h.c.})\Psi 
+a_2{\rm Tr}\overline{\Psi}\Psi(\xi M^\dagger\xi+{\rm h.c.}) 
+ a_3({\rm Tr}MU +{\rm h.c.}){\rm Tr}\overline{\Psi}\Psi \ , 
\label{eq:lagkb}
\end{eqnarray}
where the first and second terms are kinetic and mass terms of the nonlinear meson fields, $U=\exp(2i\pi_aT_a/f)$ with $\pi_a$ ($a=1\sim 8$) are the octet mesons, $T_a$ is the flavor SU(3) generator, $f$ (=93 MeV) is the meson decay constant, $\Lambda_{ {\chi} {\rm SB}}\sim$1~GeV is the chiral-symmetry breaking scale, and $M$ (=${\rm diag}(m_u, m_d, m_s)$) is the quark mass matrix. The third term in \mbox{Equation~(\ref{eq:lagkb})} is kinetic, and the mass terms of the octet baryons $\Psi$ with $M_{\rm B}$ are the spontaneously broken baryon mass. The fourth term represents the $s$-wave $K$--$B$ vector interaction with $ V_\mu\equiv 
\frac{i}{2}(\xi^\dagger\partial_\mu\xi+\xi\partial_\mu\xi^\dagger)$ being the vector current for the meson field $\xi$ (=$U^{1/2}$). This term corresponds to the Tomozawa--Weinberg term and plays a role of one of the main driving forces for KC. The fifth and sixth terms (the $F$ and $D$ terms), with $A_\mu\equiv \frac{i}{2}(\xi^\dagger\partial_\mu\xi-\xi\partial_\mu\xi^\dagger)$ being the axial-vector current for the meson, lead to the $p$-wave $K$--$B$ interactions. {As for the $p$-wave $K$--$B$ interactions, it has been suggested that a spontaneous creation of a pair of the particle--hole collective modes may occur with $K^+$ and $K^-$ quantum numbers ($p$-wave kaon condensation) in the case of a large fraction of the $\Lambda$ hyperons, through an onset mechanism that is similar to the $p$-wave pion condensation~\cite{muto2002}. In general, the $p$-wave meson condensation accompanies the particle--hole excitations of baryons, so that the onset density and the EOS with the condensed phase are sensitive to the medium effects~\cite{mtt93}, which should be taken into account for a realistic consideration. 
In this paper, we simply neglect the $F$ and $D$ terms, and only the $s$-wave KC is considered. }

The last three terms with the coefficients $a_1$$\sim$$a_3$ in \mbox{Equation~(\ref{eq:lagkb})} are in  $O(p^2)$ through the $m_q$-dependence in $M$ and break the chiral symmetry explicitly. They serve as another driving force for KC as the ``$K$-baryon sigma terms'', $\Sigma_{Kb}$. Throughout this paper, we consider only the $K^\pm$ [=($\pi_4\mp i\pi_5)/\sqrt{2}$] for $\pi_a$, and nucleons ($p$, $n$) and hyperons ($\Lambda$, $\Sigma^-$, $\Xi^-$) for $\Psi$. 

In order to reproduce the $s$-wave on-shell $K N$ scattering amplitudes, we should conventionally take into account  the range terms of the order $\omega_K^2$ [=$O(p^2)$] with the lowest kaon energy $\omega_K$ and a pole contribution from the $\Lambda$~(1405), which lies $\sim$30~MeV below the $\bar K N$ threshold. Indeed, they have sizable contributions to the $s$-wave on-shell $K N$ scattering amplitudes. 
{Nevertheless,} these contributions become negligible at a higher density $\rho_{\rm B}\gtrsim \rho_0$, since $\omega_K/m_K\ll 1$ as the density $\rho_{\rm B}$ increases, and the $\Sigma_{Kb}$ solely remains to work as the $s$-wave $K$--$B$ attractive interaction. The same {consequence}, which we call the second-order effect, has been obtained in the second-order perturbation with respect to the axial-vector current in the framework of current algebra and PCAC~\cite{fmmt1996}. Therefore, throughout this paper, these range terms and the $\Lambda$~(1405) pole contribution are neglected from the outset. 

The classical kaon field is assumed to be spatially uniform with spatial momentum $|{\bf k}|=0$ 
and represented classically as follows:
 \begin{equation}
K^\pm =\frac{f}{\sqrt{2}}\theta\exp(\pm i\mu_K t) \ , 
\label{eq:kfield}
\end{equation}
where $\theta$ is the chiral angle and $\mu_K$ is the $K^-$ chemical potential. 
{As for the $K^+$ meson in matter, the $s$-wave $K$--$N$ vector interaction and the range terms work repulsively~\cite{fmmt1996,mishra2010}. Therefore, the lowest $K^+$ energy increases with density, and the $K^+$ condensation cannot be expected to appear in $\beta$-equilibrated matter. Throughout this paper, we concentrate on $K^-$ condensation.}

By the use of Equation~(\ref{eq:kfield}), the  Lagrangian density (\ref{eq:lagkb}) is separated into the  kaon part ${\cal L}_K$ and the baryon part ${\cal L}_B$ in the mean field approximation: 
${\cal L}_{K,B}={\cal L}_K+{\cal L}_B$. 
For ${\cal L}_K$, one reads~\cite{mmt2021,muto2024}:
\begin{equation}
{\cal L}_K=f^2\Big\lbrack\frac{1}{2}(\mu_K\sin\theta)^2 - m_K^2(1-\cos\theta)
+2 \mu_K X_0 (1-\cos\theta)\Big\rbrack \ ,
\label{eq:lagk}
\end{equation}
where the second term in the bracket on the r.~h.~s. is the kaon mass term with 
\begin{equation}
m_K \equiv [\Lambda_{\chi{\rm SB}} (m_u+m_s)]^{1/2}
\label{eq:kmass}
\end{equation}
 being identified with the kaon rest mass, which is set to the empirical value (493.677~MeV). The last term in the bracket on the r.~h.~s. of Equation~(\ref{eq:lagk}) stands for the $s$-wave $K$--$B$ vector interaction, with $X_0$ being given by the following:
\begin{eqnarray}
X_0&\equiv&\frac{1}{2f^2}\sum_{b=p,n,\Lambda, \Sigma^-, \Xi^-} Q_V^b\rho_b \cr
&=& \frac{1}{2f^2}\left(\rho_p+\frac{1}{2}\rho_n-\frac{1}{2}\rho_{\Sigma^-}-\rho_{\Xi^-} \right)  \ , 
\label{eq:x0}
\end{eqnarray}
where $\rho_b$ and $Q_V^b$ 
are the number density and V-spin charge, respectively, for the baryon species $b$. The form of Equation~(\ref{eq:x0}) for $X_0$ is specified model-independently within chiral symmetry. 
From Equations~(\ref{eq:lagk}) and (\ref{eq:x0}), one can see that the $s$-wave $K$--$B$ vector interaction works attractively for protons and neutrons, but repulsively for $\Sigma^-$ and $\Xi^-$ hyperons, as far as $\mu_K > 0$, and there is no $s$-wave  $K$--$\Lambda$ vector interaction.  

For ${\cal L}_B$, one reads: 
\begin{equation}
 {\cal L}_B = \sum_{b=p,n,\Lambda, \Sigma^-, \Xi^-}\overline{\psi}_b (i\gamma^\mu \partial_\mu-M_b^\ast ) \psi_b \ , 
 \label{eq:lagb}
\end{equation} 
where $\psi_b$ is the baryon field $b$ and $M_b^\ast$ is the effective baryon mass:
\begin{equation}
M_b^\ast =M_b -\Sigma_{Kb}(1-\cos\theta) \ ,
\label{eq:effbm}
\end{equation}
where $M_b$ ($b$ = $p,n,\Lambda, \Sigma^-, \Xi^-$) is the baryon rest mass, which is read off from the last three terms in (\ref{eq:lagkb})  as follows:
\begin{eqnarray}
M_p &=& \bar M_B-2(a_1m_u+a_2m_s) \ , \cr
M_n &=& \bar M_B-2(a_1m_d+a_2m_s) \ , \cr
M_\Lambda &=&\bar M_B-1/3\cdot (a_1+a_2)(m_u+m_d+4m_s) \ , \cr
M_{\Sigma^-} &=& \bar M_B-2(a_1m_d+a_2m_u) \ , \cr
M_{\Xi^-} &=& \bar M_B-2(a_1m_s+a_2m_u)
\label{eq:brmass}
\end{eqnarray}
with $ \bar M_B=M_B - 2a_3(m_u+m_d+m_s)$. 
The quark masses $m_i$ are set to be ($m_u, m_d, m_s$) = (2.2, 4.7, 95) MeV with reference to recent results of the lattice QCD simulation~\cite{PDG2020}. 
The parameters $a_1$ and $a_2$ are then fixed to be $a_1$ = $-$0.697 and $a_2$ = 1.37 so as to reproduce the mass splittings between the octet baryons. 
The second term on the r.~h.~s. in Equation~(\ref{eq:effbm}) represents a modification of the free baryon masses  $M_b$ through the $s$-wave $K$--$B$ scalar interaction simulated by $\Sigma_{Kb}$ ($b = p, n, \Lambda, \Sigma^-, \Xi^-$), which are denoted in terms of the coefficients $a_1$, $a_2$, and $a_3$ in Equation~(\ref{eq:lagkb}) 
\begin{subequations}\label{eq:kbsigma} as follows:
\begin{eqnarray}
\Sigma_{Kn}&=&-(a_2+2a_3)(m_u+m_s) = \Sigma_{K\Sigma^-} \ ,\label{eq:kbsigma1} \\
\Sigma_{K\Lambda}&=& -\left(\frac{5}{6}a_1+\frac{5}{6}a_2+2a_3\right)(m_u+m_s) \ , \label{eq:kbsigma2} \\
\Sigma_{Kp}&=&-(a_1+a_2+2a_3)(m_u+m_s) = \Sigma_{K\Xi^-} . \label{eq:kbsigma3}
\end{eqnarray}
\end{subequations}
These quantities are identified with the ``kaon--baryon sigma terms'', which are defined by: 
\begin{equation}
\Sigma_{Kb}\equiv \frac{1}{2}(m_u+m_s)\langle b|(\bar u u +\bar s s)|b\rangle 
\label{eq:kbsigma2} 
\end{equation}
by the use of the Feynman--Hellmann theorem,  
 $ \langle b|\bar q q |b\rangle $=$\partial M_b/\partial m_q$ for $q = (u, d, s)$,
with the help of the expressions of the baryon rest masses [Equation~(\ref{eq:brmass})] up to the next-to-leading order in chiral perturbation. 
It is to be noted that, throughout this paper,  the quark content in the baryon, $ \langle b|\bar q q |b\rangle $, implies the value after the subtraction of the quark condensate in the QCD vacuum, $ \langle 0|\bar q q | 0 \rangle $~\cite{cohen1992}.
 
\section{Baryon Interactions}
\label{sec:Bforce}

\subsection{Minimal RMF for Baryon--Baryon Interaction}
\label{subsec:MRMF}

The $B$--$B$ interactions 
are given by the  exchange of mesons [$\sigma$, $\sigma^\ast$ ($\sim \bar s s$) for scalar mesons, namely 
 $\omega$, $\rho$, $\phi$ for vector mesons] in the RMF theory~\cite{mmt2021,mmt2022,muto2024}. 
 We adopt the RMF model for a two-body $B$--$B$ interaction mediated by meson exchange, without the nonlinear self-interacting (NLSI) meson potentials. [We call this model a ``minimal RMF'' (abbreviated to MRMF) throughout this paper]. 

Together with the free meson part of the Lagrangian density, one obtains the $B$--$M$ Lagrangian density as follows:
\begin{eqnarray}
{\cal L}_{B,M}&=&\sum_{b}\overline{\psi}_b \left(i\gamma^\mu D_\mu^{(b)}-\widetilde M_b^\ast \right) \psi_b \cr
&+&\frac{1}{2}\left(\partial^\mu\sigma\partial_\mu\sigma - m_\sigma^2\sigma^2\right) 
+\frac{1}{2}\left(\partial^\mu\sigma^\ast\partial_\mu\sigma^\ast-m_{\sigma^\ast}^2\sigma^{\ast 2}\right) \cr\cr
&-&\frac{1}{4}\omega^{\mu\nu}\omega_{\mu\nu}+\frac{1}{2}m_\omega^2\omega^\mu\omega_\mu 
-\frac{1}{4}R_a^{\mu\nu} R^a_{\mu\nu}+\frac{1}{2}m_\rho^2 R_a^\mu R^a_\mu \cr\cr
&-& \frac{1}{4}\phi^{\mu\nu}\phi_{\mu\nu}+\frac{1}{2}m_\phi^2\phi^\mu\phi_\mu \ ,
\label{eq:lagbm}
\end{eqnarray}
where the first term on the r.~h.~s. in Equation~(\ref{eq:lagbm}) is taken over from Equation~(\ref{eq:lagb}). 
The effective baryon mass is further modified from the $M_b^\ast$ [Equation~(\ref{eq:effbm})] due to scalar meson ($\sigma$, $\sigma^\ast$)--$B$ couplings:
\begin{eqnarray}
\widetilde M_b^\ast &\equiv& M_b^\ast - g_{\sigma b}\sigma-g_{\sigma^\ast b}\sigma^\ast \cr
&=&M_b-\Sigma_{Kb}(1-\cos\theta)- g_{\sigma b}\sigma-g_{\sigma^\ast b}\sigma^\ast \ , 
\label{eq:wtmb}
\end{eqnarray}
where $g_{\sigma b}$, $g_{\sigma^\ast b}$ are the scalar-meson--baryon coupling constants. 
{Furthermore,} the derivative in (\ref{eq:lagb}) is replaced by the covariant derivative as $\partial_\mu\rightarrow D_\mu^{(b)}\equiv \partial_\mu+i g_{\omega b}\omega_\mu+i g_{\rho b} {\hat I}_3^{\ (b)} R_\mu^3 +ig_{\phi b}\phi_\mu$, where the vector meson ($\omega$, $\rho$, $\phi$)--$B$ couplings are introduced. The vector meson fields for the $\omega$, $\rho$, $\phi$ mesons are denoted as $\omega^\mu$ and $R_a^\mu$ with the isospin components $a$ and $\phi^\mu$, respectively, and $g_{mb}$ is the vector meson--$B$ coupling constant.  The kinetic terms of the vector mesons are given in terms of 
$\omega^{\mu\nu}\equiv \partial^\mu \omega^\nu-\partial^\nu \omega^\mu$, $R_a^{\mu\nu}\equiv \partial^\mu R_a^\nu-\partial^\nu R_a^\mu$, and $\phi^{\mu\nu}\equiv \partial^\mu \phi^\nu-\partial^\nu \phi^\mu$. 
Throughout this paper, only the time components of the vector mean fields, $\omega_0$, $R_0$ ($\equiv R_0^{\rm 3}$) and $\phi_0$, are considered for the description of the ground state, and they are taken to be uniform. 
The meson masses are set to be $m_\sigma$ = 400 MeV, $m_{\sigma^\ast}$ = 975 MeV, $m_\omega$ = 783 MeV, 
$m_\rho$ = 769 MeV, and $m_\phi$ = 1020 MeV. 

 It should be remarked here that the NLSI terms were originally introduced in order to phenomenologically reproduce  the ground state properties such as the incompressibility and slope of the symmetry energy at the saturation density through $\sigma^3$ and $\sigma^4$ terms and $\omega$-$\rho$ coupling terms, respectively. The NLSI terms generate many-baryon forces through the equations of motion for the meson mean fields. The extension of the NLSI terms to high densities, however, leads to only a minor contribution to the repulsive energy. Therefore, in the context of the stiffening of the EOS at high densities, the NLSI terms are not responsible for a solution to the ``hyperon puzzle'', nor can they compensate for large attractive energy due to the appearance of the ($Y$+$K$) phase~\cite{mmt2021,mmt2022,muto2024}.

\subsection{Universal Three-Baryon Repulsive Force and Three-Nucleon Attractive Force}
\label{subsec:TBR}

Instead of the NLSI terms, we introduce the density-dependent effective two-body potentials for the universal three-baryon {repulsion} (UTBR), which has been derived from the string-junction model by Tamagaki~\cite{t2008} (SJM2) and originally applied to hyperon-mixed matter by Tamagaki, Takatsuka, and Nishizaki~\cite{tnt2008}. 
 Together with the UTBR, phenomenological three-nucleon attraction (TNA) has been taken into account, and we have obtained the baryon interaction model that reproduces the saturation properties of symmetric nuclear matter (SNM) together with empirical values of incompressibility and symmetry energy at~$\rho_0$.
 
In the SJM, when two baryons fully overlap at high densities, it is necessary to form the string junction net, accompanying the excitation of the junction ($J $)--anti-junction ($\bar J$) pair or $B$--$\bar B$ excitation with energy $\sim$2~GeV, so that the confinement mechanism persists inside two baryons. To avoid such an energy excess, two baryons are kept apart from each other, which means the existence of a high-potential core with the potential height $\sim$2~GeV between two baryons. Likewise, the origin of the three-baryon repulsive force is explained with recourse to the quark confinement mechanism in the SJM when three baryons are fully overlapped~\cite{t2008}. 
Therefore, it is natural that the three-body repulsion is qualitatively independent on the spin flavor of baryons, reflecting the confinement mechanisms of quarks at high-density regions. {Thus,} it is assumed to work universally between any baryon species. Along with this viewpoint, we adopt a specific model for the universal three-body repulsion (UTBR) proposed by Tamagaki based on the string-junction model (SJM2) ~\cite{t2008,tnt2008}.   
We utilize the density-dependent effective two-body potential $U_{\rm SJM}(1,2;\rho_{\rm B})$ between baryons 1 and 2, by integrating out variables of the third baryon participating in the UTBR, after assigning the short-range correlation function squared $f_{\rm src}(r)^2$ for each baryon pair~\cite{t2008}. 
In the following, the approximate form of $U_{\rm SJM}$ is used as follows:
\begin{equation}
U_{\rm SJM2}(r; \rho_{\rm B}) = V_r \rho_{\rm B}(1+c_r\rho_{\rm B}/\rho_0)\exp[-(r/\lambda_r)^2)] \ ,
\label{eq:aUTBR}
\end{equation}
 where $V_r$ = 95 MeV$\cdot$fm$^3$, $c_r$ = 0.024, and $\lambda_r$ = 0.86 fm corresponds to $\eta_c$ = 0.50 fm for SJM2~\cite{tnt2008}. The $U_{\rm SJM}$ grows almost linearly with $\rho_{\rm B}$. 
Finally, one obtains the effective two-body potential, 
$\widetilde U_{\rm SJM}(r;~\rho_{\rm B})~=f_{\rm src}(r) U_{\rm SJM}(r; \rho_{\rm B})$. 

To simulate the attractive contribution from the TNA to the binding energy for $\rho_{\rm B}\lesssim \rho_0$, we adopt the density-dependent effective two-body potential by Nishizaki, Takatsuka, and Hiura~\cite{nth1994}, which was phenomenologically introduced and the direct term of which agrees with the expression by Lagaris and Pandharipande~(LP1981)~\cite{lp1981}:  
\begin{equation}
U_{\rm TNA}(r; \rho_{\rm B})=V_a\rho_{\rm B} \exp(-\eta_a \rho_{\rm B})\exp[-(r/\lambda_a)^2]
(\vec{\tau}_1\cdot\vec{\tau}_2)^2 \ ,
\label{eq:tna}
\end{equation}
where the range parameter $\lambda_a$ is fixed to be 2.0 fm. 
 $U_{\rm TNA}(r; \rho_{\rm B})$ depends upon not only on density but also on 
 isospin $\vec \tau_1\cdot\vec \tau_2$ with Pauli matrices $\vec \tau_i$. The parameters $V_a$ and $\eta_a$ are determined together with 
other parameters to reproduce the saturation properties of the SNM for the allowable values of $L$ (see Section~\ref{sec:SNM}). 

\section{Description of the Ground State for the ($Y$+$K$) Phase}
\label{sec:ground-state}

\subsection{Energy Density Expression for the ($Y$+$K$) Phase}
\label{subsec:energy}

The energy density ${\cal E}$ for the ($Y$+$K$) phase is separated into the KC part, ${\cal E}_K$, the baryon kinetic part, and meson part for two-body baryon interactions, ${\cal E}_{B,M}$, three-body interaction parts, ${\cal E}$~(UTBR) + ${\cal E}$~(TNA), and free lepton parts, ${\cal E}_e$ for the ultra-relativistic electrons and ${\cal E}_\mu$ for muons. 
From (\ref{eq:lagk}) and (\ref{eq:lagbm}) one obtains the following:
\begin{equation}
{\cal E}_K=\frac{1}{2}(\mu_K f\sin\theta)^2+f^2m_K^2(1-\cos\theta) \ , 
\label{eq:ekfinal}
\end{equation}
\vspace{-0.5cm}~
\begin{eqnarray}
{\cal E}_{B,M}&=&\sum_b \frac{2}{(2\pi)^3}\int_{|{\bf p}|\leq p_F(b)} d^3|{\bf p}|(|{\bf p}|^2+\widetilde M_b^{\ast 2})^{1/2} \cr
&+&\frac{1}{2}\left(m_\sigma^2\sigma^2+m_{\sigma^\ast}^2\sigma^{\ast 2}\right) 
+ \frac{1}{2}\left(m_\omega^2\omega_0^2+m_\rho^2 R_0^2+m_\phi^2\phi_0^2\right) \ ,  
\label{eq:ebm}
\end{eqnarray}
where {each Fermi sphere of a baryon ($b$) is fully occupied up to} Fermi momentum $p_F(b)$. 
 
The contribution from the UTBR is written in the Hartree approximation as follows:
\begin{eqnarray}
{\cal E}~({\rm UTBR}) &=& 2\pi\rho_{\rm B}^2\int d rr^2 \widetilde U_{\rm SJM2}(r; \rho_{\rm B}) \cr
&=& \frac{\pi^{3/2}}{2}V_r (\widetilde\lambda_r)^3\rho_{\rm B}^3\left(1+c_r\frac{\rho_{\rm B}}{\rho_0}\right) \cr
&=&  \frac{\pi^{3/2}}{2}\rho_{\rm B}^2 U_{\rm SJM2}(r=0; \rho_{\rm B})\cdot (\widetilde\lambda_r)^3 \ ,  
\label{eq:edUTBRtil}
\end{eqnarray}
where $\displaystyle (\widetilde\lambda_r)^3\equiv \frac{4}{\pi^{1/2}}\int_0^\infty dr r^2f_{\rm src}(r)e^{-(r/\lambda_r)^2}$ (=0.589496 $\cdots$) for SJM2. With the use of the spatial average for the s.~r.~c. function $f_{\rm src}(r)$ being denoted as $\overline{f}_{\rm src}$, one can write $(\widetilde\lambda_r)^3\simeq \overline{f}_{\rm src}\cdot (\lambda_r)^3$. {Thus,} $\widetilde\lambda_r$ is interpreted as the range of the effective two-body potential $\widetilde U_{\rm SJM2}(r; \rho_{\rm B})$.

Likewise, the energy density contribution from the direct term of the TNA is represented as follows:
\begin{eqnarray}
{\cal E}~({\rm TNA})&=&\frac{1}{2} \int d^3 r V_a\rho_{\rm B} e^{-\eta_a \rho_{\rm B}} e^{-(r/\lambda_a)^2} \cr
&\times&\rho_{\rm B}^2\lbrace 3-2(1-2x_p)^2\rbrace \cr
&=&\gamma_a\rho_{\rm B}^3e^{-\eta_a\rho_{\rm B}}\lbrace 3-2(1-2x_p)^2\rbrace 
\label{eq:edTNA}
\end{eqnarray}
with $\displaystyle\gamma_a\equiv(\pi^{3/2}/2) V_a\lambda_a^3$ and $x_p=\rho_p/\rho_{\rm B}$ being the proton-mixing ratio. 
With Equations~(\ref{eq:ekfinal})$-$(\ref{eq:edTNA}) and the relativistic forms of the lepton energy densities, the total energy density ${\cal E}$ is given~by the following:
\begin{equation}
{\cal E}={\cal E}_K+{\cal E}_{B,M}+{\cal E}({\rm UTBR})+{\cal E}({\rm TNA})+{\cal E}_e+{\cal E}_\mu \ .
\label{eq:total-edensity}
\end{equation} 

\subsection{Classical Field Equations for Kaon Condensates and Meson Mean Fields}
\label{subsec:eom}

Throughout this paper, the classical $K^-$ field ($\displaystyle |K^-|=f\theta/\sqrt{2}$) 
and meson mean fields ($\sigma, \sigma^\ast, \omega, \rho, \phi$) 
are set to be uniform and only depend on total baryon density $\rho_{\rm B}$. 
The equations of motion for these fields are derived from the Lagrangian density ${\cal L}_K+{\cal L}_{B,M}$ in the mean field approximation. 

The classical kaon field equation follows from 
\[ \partial({\cal L}_K+{\cal L}_{B,M})/\partial\theta=0 \ ,  \] 
which renders
\begin{equation}
\mu_K^2\cos\theta+2X_0\mu_K-m_K^{\ast 2}=0 \ , 
\label{eq:keom2}
\end{equation} 
where the effective kaon mass squared is defined by
\begin{equation}
m_K^{\ast 2}\equiv m_K^2-\frac{1}{f^2}\sum_{b=p,n,\Lambda, \Sigma^-, \Xi^-}\rho_b^s\Sigma_{Kb}  
\label{eq:ekm2}
\end{equation}
with $\rho_b^s$ being a scalar density for baryon $b$: 
\begin{equation}
 \rho_b^s=\frac{2}{(2\pi)^3} \int_{|{\bf p}|\leq p_F(b)} d^3|{\bf p}|\frac{\widetilde M_b^\ast}{(|{\bf p}|^2
+\widetilde M_b^{\ast 2})^{1/2}} \ . 
\label{eq:rhobs}
 \end{equation}
As for the equations of motion for meson {mean} fields, one obtains the following:
\begin{subequations}\label{eq:cieom}
\begin{eqnarray}
m_\sigma^2\sigma&=&\sum_{b=p,n,\Lambda,\Sigma^-,\Xi^-}g_{\sigma b}\rho_b^s \ , \label{eq:cieom1} \\
m_\sigma^{\ast 2}\sigma^\ast&=&\sum_{Y=\Lambda,\Sigma^-,\Xi^-}g_{\sigma^\ast Y}\rho_Y^s \ , \label{eq:cieom2}\\
m_\omega^2\omega_0&=&\sum_{b=p,n,\Lambda,\Sigma^-,\Xi^-} g_{\omega b}\rho_b \ ,  \label{eq:cieom4}\\
m_\rho^2 R_0&=&\sum_{b=p,n,\Lambda,\Sigma^-,\Xi^-} g_{\rho b}{\hat I}_3^{(b)} \rho_b \ , \label{eq:cieom5} \\
m_\phi^2\phi_0 &=&\sum_{Y=\Lambda,\Sigma^-,\Xi^-}g_{\phi Y}\rho_Y \ . \label{eq:cieom6}
\end{eqnarray}
\end{subequations}

\subsection{Ground State Conditions}
\label{subsec:grcond}

The ground state energy for the ($Y+K$) phase is obtained under the charge neutrality, baryon number, and $\beta$-equilibrium conditions. 
{Since we consider $K^-$ condensation in hyperon-mixed matter, and only $p$, $n$, $\Sigma^-$, and $\Xi^-$ are taken into account for baryons, as stated in Section~\ref{sec:kaon}, the} charge neutrality condition is written as  follows:
\begin{equation}
\rho_Q=\rho_p-\rho_{\Sigma^-}-\rho_{\Xi^-}-\rho_{K^-}-\rho_e-\rho_\mu=0 \ , 
\label{eq:charge}
\end{equation}
where $\rho_Q$ denotes the total negative charge density, and $\rho_{K^-}$ is the number density of KC and is given from the kaon part of the Lagrangian density (\ref{eq:lagk}) as:
\begin{eqnarray}
\rho_{K^-}&=&-iK^-(\partial{\cal L}_K/\partial\dot{K^-})+iK^+(\partial{\cal L}_K/\partial\dot{K^+}) \cr
&=&\mu_K f^2\sin^2\theta+2f^2X_0(1-\cos\theta) \ . 
\label{eq:rhokc}
\end{eqnarray}
In Equation~(\ref{eq:charge}), $\rho_e$ is the electron number density and is related to the electron chemical potential $\mu_e$ as $\rho_e=\mu_e^3/(3\pi^2)$ in the ultra-relativistic limit. $\rho_\mu$ is the muon number density and is given by $\rho_\mu=[p_F(\mu^-)]^3/(3\pi^2)$. 

The baryon number conservation is given by:
\begin{equation}
\rho_p +\rho_n + \rho_\Lambda+\rho_{\Sigma^-}+\rho_{\Xi^-}=\rho_{\rm B} \ .
\label{eq:bn}
\end{equation}
In addition, the following chemical equilibrium conditions for weak processes are imposed: 
 $n\rightleftharpoons p+K^-$, $n\rightleftharpoons p+e^- (+\bar\nu_e)$, $n + e^-\rightleftharpoons \Sigma^-(+\nu_e)$, $\Lambda + e^-\rightleftharpoons \Xi^-(+\nu_e)$, $n\rightleftharpoons \Lambda(+\nu_e\bar\nu_e)$, and those involved in muons in place of $e^-$ if muons are present. 
 These conditions are followed by the relations between the chemical potentials 
\begin{eqnarray}
\mu=\mu_K&=&\mu_e=\mu_\mu=\mu_n-\mu_p \ , \cr
\mu_\Lambda&=&\mu_n , \cr
 \mu_{\Sigma^-}&=&\mu_{\Xi^-}=\mu_n+\mu_e \ ,
\label{eq:chem}
\end{eqnarray}
where $\mu$ and $\mu_i$ (=$\partial{\cal E}/\partial \rho_i$) ($i$ = $p$, $n$, $\Lambda$, $\Sigma^-$, $\Xi^-$, $K^-$, $e^-$, $\mu^-$) are the charge chemical 
potential and the chemical potential for each particle species ($i$), respectively, at a given baryon number density $\rho_{\rm B}$. 
{It is to be noted that the strangeness-changing weak process, $n\rightarrow p+K^-$, is expressed in terms of quarks as $d\rightarrow u+ W^-$ and $W^- \rightarrow s +\bar u$, the latter of which proceeds through a flavor-mixing effect with the Cabibbo suppression.}

\section{{Choice of Parameters and} Properties of Symmetric Nuclear Matter}
\label{sec:SNM}

\subsection{Meson--Nucleon Coupling Constants Determined from Saturation Properties in the SNM}
\label{subsec:saturation}

In order to determine the meson--nucleon coupling constants, $g_{\sigma N}$, $g_{\omega N}$, $g_{\rho N}$, and the $\sigma$, $\omega$ mean fields, $\langle\sigma\rangle_0$, $\langle\omega_0\rangle_0$, and parameters in TNA, $\eta_a$, $\gamma_a$, we impose the saturation properties of the symmetric nuclear matter (SNM), i.e., the saturation density $\rho_0$ = 0.16 fm$^{-3}$, binding energy $B_0$ = 16.3 MeV,  incompressibility $K$ = 240 MeV, symmetry energy $S_0$ = 31.5 MeV, and  slope $L$ [$\equiv 3\rho_0\left(d S(\rho_{\rm B}) /d \rho_B\right)_{\rho_B=\rho_0} $] = (60--70) MeV, 
taking into account the ambiguity of the empirical value of the $L$~\cite{oertel2017}. Also, the equations of motion for the meson mean fields are imposed:  
\begin{eqnarray}
m_\sigma^2\langle\sigma\rangle_0&=&g_{\sigma N}\rho_N^s\vert_{\rho_{\rm B}=\rho_0} \cr
m_\omega^2\langle\omega_0\rangle_0&=&g_{\omega N}\rho_0 \ , 
\label{eq:eomSNM}
\end{eqnarray}
where $\rho_N^s$ (=$\rho_p^s+\rho_n^s$) is the nuclear scalar density. 
In Table~\ref{tab:para1}, the relevant quantities associated with the (MRMF+UTBR+TNA) model are listed for three cases of $L$ = (60, 65, 70) MeV. 
\begin{table}[h]
\caption{The parameters $\gamma_a$, $\eta_a$ for TNA, the coupling constants, $g_{\sigma N}$, $g_{\omega N}$, $g_{\rho N}$, the meson mean fields, $\langle\sigma\rangle_0$, $\langle\omega_0\rangle_0$, and the effective mass ratio for the nucleon, 
$(M_N^\ast/M_N)_0$, in the SNM at $\rho_{\rm B}$ = $\rho_0$, obtained in the (MRMF+UTBR+TNA) model
 in cases of $L$ = 60, 65, and 70 MeV. The  $\sigma$--$Y$ coupling constants ($Y$=$\Lambda$, $\Sigma^-$, $\Xi^-$) determined from the potential depths for $Y$ in the SNM are also listed.}
\begin{tabular}{ c || c | c | c | c | c || c | c || c || c | c | c }
\hline
 & $\gamma_a$ & $\eta_a$  &  $g_{\sigma N}$ & $g_{\omega N}$  & $g_{\rho N}$ & $\langle\sigma\rangle_0$ & $\langle\omega_0\rangle_0$  & $(M_N^\ast/M_N)_0$ & $g_{\sigma\Lambda}$ & $g_{\sigma\Sigma^-}$ & $g_{\sigma\Xi^-}$ \\ 
  & (MeV$\cdot$fm$^6$) & (fm$^3$) &          &         &   &  (MeV) &  (MeV)  &  &  &  &  \\     \hline\hline
SJM2+TNA-L60    & $-$1662.63     &  17.18 &  5.27 & 8.16  & 3.29 & 39.06 & 16.37  & 0.78 & 3.29 & 2.00 & 1.82  \\
SJM2+TNA-L65    & $-$1597.67     &  18.25 &  5.71 & 9.07  & 3.35 & 42.16 & 18.18  & 0.74 & 3.54 & 2.34 & 1.93   \\
SJM2+TNA-L70    & $-$1585.48     &  19.82 &  6.07 & 9.77  & 3.41 & 44.62 & 19.59  & 0.71 & 3.74 & 2.61 & 2.02   \\
\hline
\end{tabular}
\label{tab:para1}
\end{table}
One can see from Table~\ref{tab:para1} that the slope $L$ sensitively affects  both $\sigma$ and $\omega$ mean field values at the saturation density $\rho_0$; these meson mean fields contribute to the binding energy at $\rho_0$ by adjusting to the change in attractive energy contribution from the TNA due to the change in the $L$. The change in these meson mean fields prevails at high densities beyond $\rho_0$ and affects the stiffness of the EOS at high densities. See also Section~\ref{sec:MR}. For the choice of the $L$ within the (MRMF+UTBR+TNA) model, we refer to Refs.~\cite{mmt2021,muto2024}. 

{Both} the TNR and TNA play important roles in locating the total energy minimum at the empirical saturation point. Indeed, it is necessary to include both the TNR and the TNA in the total energy $E$ (total) in addition to the nuclear two-body interaction within the MRMF, in order to reproduce the empirical saturation property and incompressibility (=240 MeV) for the SNM. The TNR  (the TNA) pushes up (pulls down) the $E$ (two-body) curve for $\rho_{\rm B}\gtrsim \rho_0$ ($\rho_{\rm B}\lesssim \rho_0$). 

\subsection{{Meson--Hyperon Coupling Constants}}
\label{subsec:mY-coupling}

The meson--hyperon coupling constants are determined to obtain the hyperon--nucleon and hyperon--hyperon interactions.
The vector meson couplings for hyperons are related to the vector meson--nucleon couplings $g_{\omega N}$, $g_{\rho N}$, $g_{\phi N}$ through the SU(6) symmetry relations~\cite{sdg94}: 
 \begin{subequations}
\begin{eqnarray}\label{eq:gmY}
g_{\omega\Lambda}&=&g_{\omega\Sigma^-}=2g_{\omega \Xi^-}=(2/3) g_{\omega N} \ , 
\label{eq:gmY1} \\
g_{\rho \Lambda}&=& 0 \ , g_{\rho\Sigma^-}=2g_{\rho\Xi^-}=2g_{\rho N} \ , \label{eq:gmY2} \\ 
 g_{\phi\Lambda}&=& g_{\phi\Sigma^-}=(1/2) g_{\phi\Xi^-}=-(\sqrt{2}/3) g_{\omega N} \ . 
\end{eqnarray}
\end{subequations}
The scalar ($\sigma$, $\sigma^\ast$) meson--hyperon couplings are determined with the help of information  from the phenomenological analyses of recent hypernuclear experiments. 
The obtained values of the $\sigma$-$Y$ coupling constants, $g_{\sigma Y}$, are listed for the cases of $L$ = (60, 65, 70)~MeV in Table~\ref{tab:para1} in Section~\ref{subsec:saturation}. 
The details of obtaining the $g_{\sigma Y}$ and 
the $\sigma^\ast$-$Y$ coupling constants, $g_{\sigma^\ast Y}$, are addressed in Ref.~\cite{mmt2022}.

\section{Estimation of the kaon--baryon Sigma Terms --- Quark Contents in the~Baryon}
\label{sec:qcb}

\subsection*{Nonlinear Effect on the Quark {Contents}}
\label{subsec:NLQC}

In this section, we estimate the allowable value of the $K$-nucleon sigma term, $\Sigma_{KN}$ ($N = p,n$). 
$\Sigma_{KN}$ is generally 
{expressed} as:
\begin{subequations}\label{eq:knsigma2}
\begin{eqnarray}
\Sigma_{KN}&=&\frac{m_u+m_s}{2\hat m}\left(\frac{\Sigma_{\pi N}}{1+z_N}+\frac{\hat m}{m_s}\sigma_s\right)  \label{eq:knsigma2-1} \\
&=&\frac{m_u+m_s}{2\hat m}\Sigma_{\pi N}\left(\frac{1}{1+z_N}+\frac{1}{2}y_N\right) 
\label{eq:knsigma2-2}
\end{eqnarray}
\end{subequations}
with $\hat m\equiv (m_u+m_d)/2$. 
In Equation~(\ref{eq:knsigma2-1}), $\Sigma_{\pi N}$ is the $\pi N$ sigma term,  
\begin{equation}
\Sigma_{\pi N} \equiv \hat m \langle N|(\bar uu+\bar dd)|N\rangle \ ,
\label{eq:sigma-piN}
\end{equation}
and $\sigma_s$ ($\equiv m_s\langle N|\bar s s |N\rangle$) is the the strangeness {content} in the nucleon. In Equation~(\ref{eq:knsigma2}b), $z_N$ and $y_N$ are defined by
\begin{equation}
z_N\equiv \langle N|\bar d d|N\rangle / \langle N|\bar u u|N\rangle \ , 
\label{eq:zN}
\end{equation}
\begin{equation}
y_N\equiv 2\langle N| \bar s s|N\rangle/\langle N|(\bar u u+\bar d d) |N\rangle \ .
\end{equation} 
The former stands for the isospin asymmetry for the quark {content} in the nucleon, and the latter implies breaking scale of the Okubo--Zweig--Iizuka (OZI) rule. 
The $KN$ sigma terms are related to the following flavor nonsinglet {contents} as well:
\begin{subequations}\label{eq:sigma03}
\begin{eqnarray}
\sigma_0&\equiv&\hat m\langle N|(\bar u u + \bar d d-2\bar s s)|N\rangle = \Sigma_{\pi N}-(2\hat m/m_s) \sigma_s \ , \label{eq:sigma0} \\
\sigma_3&\equiv&\hat m \langle p|(\bar u u -\bar d d )|p\rangle \ .
\label{eq:sigma3}
\end{eqnarray}
\end{subequations}
In chiral perturbation theory, these {contents} are related to the mass difference between the octet baryons:
\begin{subequations}\label{eq:psigma03}
\begin{eqnarray}
\sigma_0&=&-2\hat m(a_1-2a_2) \simeq \frac{3}{m_s/\hat m-1}~(M_{\Xi}-M_\Lambda) \simeq 25~{\rm MeV} \ , \label{eq:psigma0} \\
\sigma_3 &=& -2\hat m a_1\simeq \frac{1}{m_s/\hat m -1}(M_{\Xi}-M_{\Sigma})\simeq 5~{\rm  MeV} \label{eq:psigma3}
\end{eqnarray}
\end{subequations}
with the use of Equation~(\ref{eq:brmass}). 
Recent lattice QCD results suggest small $\bar s s $ {contents} in the nucleon, i.e., $y_N\equiv 2\langle N| \bar s s|N\rangle/\langle N|(\bar u u+\bar d d) |N\rangle$ = 0.03--0.2~\cite{ohki08,Durr2016,Alex2020}. 
In particular, for $\sigma_s\sim 0$, one can see from Equations~(\ref{eq:sigma0}) and (\ref{eq:psigma0}) 
that $\sigma_0=\Sigma_{\pi N}\simeq$ 25~MeV. This value of $\Sigma_{\pi N}$ is too small as compared with the phenomenological values (40--60) MeV~\cite{gls1991,a2021}, which are deduced from the analyses of $\pi$-$N$ scattering and pionic atoms, or lattice QCD results \mbox{$\sim$40 MeV}~\cite{a2021}.
{Thus,} as far as the estimation of the quark {contents} in the nucleon is based on chiral perturbation, small $\bar s s$ {contents} in the nucleon are incompatible with the standard value of the $\pi N$ sigma term. 

It has been shown that nonlinear effects beyond chiral perturbation can make both the value of $\Sigma_{\pi N}$ and the octet baryon mass splittings
 consistent with experiments with a small strangeness content of the proton~\cite{a2021,jk1987,hk1991}.  
Here, we take into account the nonlinear effect on the strangeness quark {content}, which originates from the additional universal rest mass contribution of baryons, $\Delta M(m_s)$, in a higher order with respect to $m_s$.

The $\bar q q$ contents in the baryon $b$ after correction from the nonlinear effect are obtained from $ \langle b|\bar q q |b\rangle $=$\partial \widetilde M_b/\partial m_q$, with the baryon rest masses given by $\widetilde M_b = M_b+\Delta M(m_s)$. The result is:
\begin{eqnarray}
 \langle p|\bar u u |p\rangle &=&  \langle n|\bar d d |n\rangle = - 2(a_1+a_3) \ , \cr
  \langle p|\bar d d |p\rangle &=&  \langle n|\bar u u |n\rangle = - 2a_3 \ , \cr
    \langle p|\bar s s |p\rangle &=&  \langle n|\bar s s |n\rangle = - 2(a_2+a_3)+\Delta \ 
    \label{eq:qNcontent}
\end{eqnarray}
with $\Delta\equiv d \Delta M(m_s)/dm_s$, and
\begin{eqnarray}
 \langle \Lambda|\bar u u |\Lambda\rangle &=&  \langle \Lambda|\bar d d |\Lambda\rangle = - \frac{1}{3}(a_1+a_2)-2a_3 \ , \cr
  \langle \Lambda|\bar s s |\Lambda\rangle &=& - \frac{4}{3}(a_1+a_2)-2a_3 +\Delta \ , \cr
\langle \Sigma^-|\bar u u  |\Sigma^- \rangle &=& -2(a_2+a_3) \ , \langle \Sigma^-|\bar d d  |\Sigma^- \rangle = -2(a_1+a_3) \ , \cr
  \langle \Sigma^-|\bar s s |\Sigma^-\rangle &=& - 2a_3 +\Delta \ , \cr
\langle \Xi^-|\bar u u  |\Xi^- \rangle &=& -2(a_2+a_3) \ , \langle \Xi^-|\bar d d  |\Xi^- \rangle = -2a_3 \ , \cr
  \langle \Xi^-|\bar s s |\Xi^-\rangle &=& - 2(a_1+a_3)+\Delta \ .
    \label{eq:qYcontent}
\end{eqnarray}
Then, the $Kb$ sigma terms (\ref{eq:kbsigma}) are modified by the replacement: $a_3\rightarrow \widetilde a_3\equiv a_3-\Delta/4$.   
The nonlinear effect $\Delta$ is absorbed into $\widetilde a_3$. 

The ``$K$--baryon sigma terms'' $\Sigma_{Kb}$ are
 given in terms of $a_1\sim a_3$, $\Delta$, and $m_u$, $m_d$, $m_s$ by
\begin{subequations}\label{eq:Akbsigma}
\begin{eqnarray}
\Sigma_{Kn}&=&-(a_2+2\widetilde a_3)(m_u+m_s) = \Sigma_{K\Sigma^-} \ ,\label{eq:Akbsigma1} \\
\Sigma_{K\Lambda}&=& -\left(\frac{5}{6}a_1+\frac{5}{6}a_2+2\widetilde a_3\right)(m_u+m_s) \ , \label{eq:Akbsigma2} \\
\Sigma_{Kp}&=&-(a_1+a_2+2\widetilde a_3)(m_u+m_s) = \Sigma_{K\Xi^-}  \label{eq:Akbsigma3}
\end{eqnarray}
\end{subequations}
with $\widetilde a_3\equiv a_3-\Delta/4$. 

\begin{figure}[h]
\includegraphics[height=0.30\textheight]{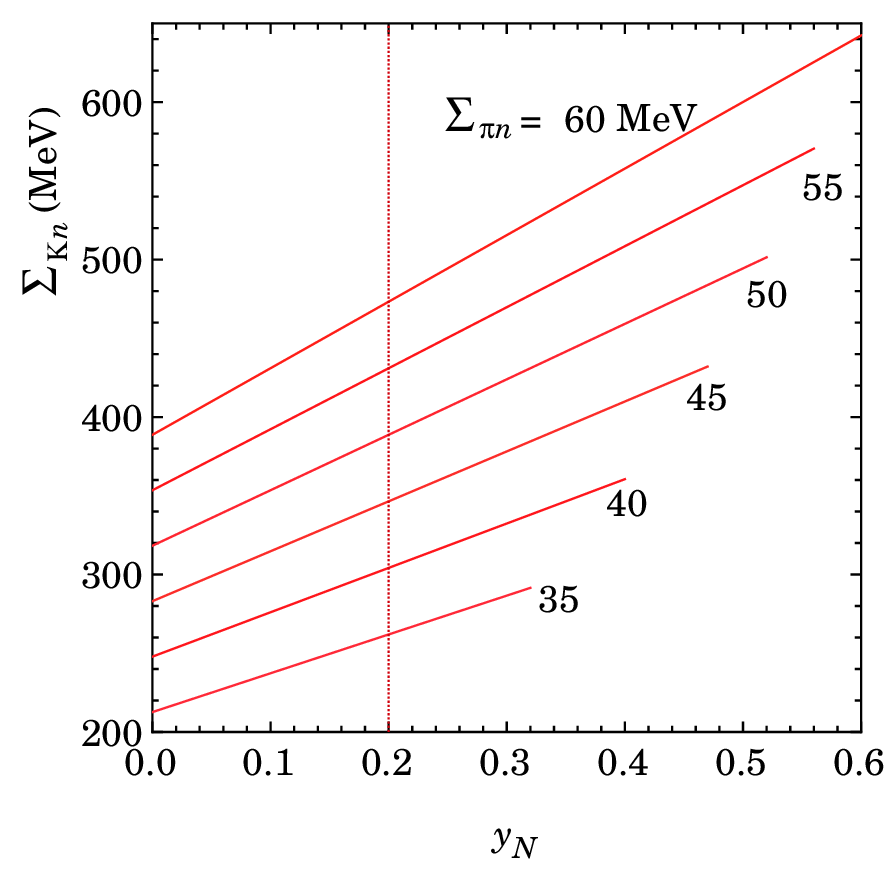}~
\caption{{The} $K$-neutron sigma term $\Sigma_{Kn}$ as a function of $y_N$ ($\equiv 2\langle N| \bar s s|N\rangle/\langle N|(\bar u u+\bar d d) |N\rangle$) for a given value of $\Sigma_{\pi N}$ = (35--60)~MeV. The current quark masses are set to ($m_u, m_d, m_s$) = (2.2, 4.7, 95)~MeV. The vertical dotted line denotes the upper value of $y_N$ = 0.2 suggested by the lattice QCD results, taken from~\cite{ohki08,Durr2016, Alex2020}. The right endpoint of each line corresponds to $\Delta$ = 0 (the case of chiral perturbation). See the text for details. }
\label{fig:sigkn}
\end{figure}

The parameters $z_N$ and $y_N$ are rewritten specifically as 
\begin{equation}
z_p=\frac{a_3}{a_1+a_3}=1/z_n \ ,  
\label{eq:zN1} 
\end{equation}
\begin{equation}
y_N =\frac{2(a_2+ a_3)-\Delta}{a_1+2 a_3} \qquad (N=p, n) \ . 
\label{eq:yN}
\end{equation}
Once $\Sigma_{\pi N}$ is given, $a_3$ and $z_p$ ($z_n$) are determined from Equations~(\ref{eq:sigma-piN}) and (\ref{eq:zN1}), respectively. 
Then, $\Sigma_{KN}$ is obtained from Equation~(\ref{eq:knsigma2-2}) together with $y_N$, which is given as a function of $\Delta$ through Equation~(\ref{eq:yN}). 
With the nonlinear effect $\Delta$, $\sigma_0$ is represented as $\sigma_0=-2\hat m (a_1-2a_2+\Delta)$. 
In Figure~\ref{fig:sigkn}, the $K$-neutron sigma term $\Sigma_{Kn}$ as a function of $y_N$ is shown at a fixed value of $\Sigma_{\pi N}$ considering uncertainty ranging from  35~MeV to 60~MeV. The vertical dotted line shows boundaries of the allowable region for $y_N$, taken from~\cite{ohki08,Durr2016, Alex2020}.
The standard value for $\Sigma_{\pi N}$ has been taken to be $\sim$ 45~MeV phenomenologically~\cite{gls1991}. Recently, the higher values (50--60~MeV) were suggested from the phenomenological analyses of $\pi$-$N$ scatterings~\cite{a2021}. 
In view of this, reading off from Figure~\ref{fig:sigkn}, we take two cases of $\Sigma_{Kn}$ = 300 MeV with $y_N$ = 0 and 400 MeV with $y_N$ = 0.2 as typical values for $\Sigma_{Kn}$ throughout this paper. The corresponding quantities, $\Sigma_{\pi N}$, $\Delta$, and $\Sigma_{Kb}$ ($b=p, \Lambda, \Sigma^-, \Xi^-$) together with $a_3$, $\widetilde a_3$ are also determined for fixed values of ($m_u$, $m_d$, $m_s$) and ($a_1$, $a_2$). 
The results are listed in Table~\ref{tab:kbsigma}. 
\begin{table}[h]
\caption{The parameters $a_1$, $a_2$, $a_3$, and $\widetilde a_3$ in the chiral symmetry breaking terms in the effective chiral Lagrangian (\ref{eq:lagkb}), and the quantities in terms of them for the current quark masses ($m_u$, $m_d$, $m_s$) = (2.2, 4.7, 95)~MeV~\cite{PDG2020}: $y_N\equiv 2\langle N| \bar s s|N\rangle/\langle N|(\bar u u+\bar d d) |N\rangle$, $\Delta$ being a shift of the strangeness content in the nucleon from the value in the leading-order chiral perturbation, $\Sigma_{\pi N}$ the $\pi N$ sigma term, and the ``K--Baryon sigma terms'' $\Sigma_{Kb}$ ($b=p, n, \Lambda, \Sigma^-, \Xi^-$) adopted in this work. The $K$-neutron sigma term, $\Sigma_{Kn}$, is set to be the two typical values 300 MeV and 400 MeV. The $K^-$ optical potential $U_K$ in the SNM is listed for each value of $\Sigma_{Kn}$. 
}
\begin{center}
\begin{tabular}{c || c  c || c c | c c c c ||c }
\hline
 ($a_1, a_2$) & $a_3$ & $\Sigma_{\pi N}$ & $y_N$ & $\Delta$ & $\widetilde a_3$ & $\Sigma_{Kn}$ (= $\Sigma_{K\Sigma^-}$)   &  $\Sigma_{Kp}$ (= $\Sigma_{K\Xi^-}$)  & $\Sigma_{K\Lambda}$ & $U_K$ \\
 & & (MeV)  & & &  & (MeV) & (MeV) & (MeV)  & (MeV) \\ \hline\hline
  ($-$0.697, 1.37) & $-$3.09 & 47.4  & 0 &$-$3.43 & $-$2.23 & 300 & 368 & 379  & $-$111 \\
  & $-$3.37 & 51.3 & 0.20 & $-$2.51 & $-$2.74  & 400 & 468 & 479 & $-$131 \\\hline
\end{tabular}
\label{tab:kbsigma}
\end{center}
\end{table} 

A scale of the $s$-wave $K$--$N$ attraction is characterized by the $K^-$ optical potential $U_K$ in the SNM, which is defined in terms of the $K^-$ self-energy [Equation~(\ref{eq:selfk2}) in Section~\ref{subsec:onsetKC}] as $U_K=\Pi_K(\omega_K; \rho_{\rm B}) /(2\omega_K)\vert_{\rho_{\rm B}=\rho_0}$. In Table~\ref{tab:kbsigma}, the $U_K$ is listed for each case of $\Sigma_{Kn}$. 
The value of $U_K$ has a sensitive dependence on $\Sigma_{Kn}$, while it depends little on the slope $L$. 
Our deduced value of the depth $|U_K|$ (110~MeV--130~MeV) is larger than the theoretical values in the chiral unitary approach~\cite{ro2000}, while it is similar to that of Refs.~\cite{wrw1997,ww1997} ($U_K \sim -$120~MeV) with the inclusion of short-range correlations. It can also be compared with the recent optimal value of the real part of the $K^-$ optical potential depth $|V_0|$ = 80 MeV with the imaginary part $W_0$ = $-$40 MeV in the J-PARC~E05 experiment~\cite{ichikawa2020}. 

It is to be noted that the pion-baryon sigma terms [$\sigma_{bq}=\hat m \langle b|(\bar u u +\bar d d)|b\rangle$] and strangeness sigma terms [$\sigma_{bs}=m_s \langle b|\bar s s|b\rangle$] in the octet baryons ($b$) have been derived from analyses of the lattice QCD simulations for the octet baryon masses~\cite{sty2013,Lutz2014}. 
For comparison, we estimate $\bar \sigma_{bq}\equiv\hat m\langle b|(\bar u u +\bar d d)|b\rangle/M_b$ and $\bar\sigma_{bs}\equiv m_s \langle b|\bar s s|b\rangle/M_b$ with $M_b$ being the empirical baryon mass. For $\Sigma_{\pi N}$=45~MeV and $y_N=0.04$, which are referred to from~\cite{sty2013}, one obtains
$\Delta$ = $-$2.82 ($a_3$ = $-$2.91), $\bar \sigma_{bq}$ = (0.024, 0.017, 0.015, 0.012),
and $\bar\sigma_{bs}$ = (0.026, 0.179, 0.238, 0.316) for $b$ = ($N$, $Y$ (=$\Lambda$, $\Sigma^-$, $\Xi^-$)). 
The relative ordering of $\bar\sigma_{bq}$, $\bar\sigma_{bs}$ for $b= (N, Y(=\Lambda, \Sigma^-, \Xi^-))$ estimated in our model agrees well with those in~\cite{sty2013}, while there is a little difference in the absolute values for $\sigma_{Yq}$, $\sigma_{Ys}$ between our results and those in Figure~2 in~\cite{sty2013}.

\section{Onset of KC and Composition of Matter in the ($Y$+$K$) Phase}
\label{sec:KP}

Here, we consider kaon properties in hyperon-mixed matter and obtain the onset density of KC with our interaction model (ChL+MRMF+UTBR+TNA). 

\subsection{Onset Density of Kaon Condensation in Hyperon-Mixed Matter}
\label{subsec:onsetKC}

The in-medium modification of kaon dynamics in dense matter is revealed by the density dependence of the lowest kaon energy $\omega_K(\rho_B)$. 
 $\omega_K(\rho_B)$ is given as a pole of the kaon propagator at $\rho_B$, i.e., $D_K^{-1}(\omega_K; \rho_{B})=0$. The kaon inverse propagator, $D_K^{-1}(\omega_K; \rho_{B})$, is obtained through the expansion of the effective energy density with respect to the classical kaon field, 
\begin{equation}
{\cal E}_{\rm eff}(\theta)={\cal E}_{\rm eff}(0)-\frac{f^2}{2}D_K^{-1}(\mu; \rho_{\rm B})\theta^2+O(\theta^4)  \ ,
\label{eq:dkinv1}
\end{equation}
where ${\cal E}_{\rm eff}\equiv{\cal E}+\mu\rho_Q+\nu\rho_{\rm B}$ with the baryon chemical potential $\nu$. 
With the use of \mbox{Equations~(\ref{eq:ekfinal}), (\ref{eq:ebm}), (\ref{eq:charge}), and (\ref{eq:rhokc})}, and 
by setting $\mu_K\rightarrow \omega_K$, $\theta\rightarrow 0$, one obtains the following:
\begin{equation}
D_K^{-1}(\omega_K; \rho_{B})
=\omega_K^2-m_K^2-\Pi_K(\omega_K; \rho_B) \ , 
\label{eq:dkinv}
\end{equation}
where $\Pi_K(\omega_K; \rho_B)$ is the kaon self-energy: 
\begin{equation}
\Pi_K(\omega_K; \rho_B)
 = -\frac{1}{f^2}\sum_{b=p,n,\Lambda, \Sigma^-, \Xi^-}\left(\rho_b^s\Sigma_{Kb}+\omega_K\rho_bQ_V^b\right) \ . 
\label{eq:selfk2}
\end{equation}
From Equations~(\ref{eq:dkinv}) and (\ref{eq:selfk2}), the $\omega_K$ is given explicitly as:
\begin{equation}
\omega_K=-X_0+\left(X_0^2+m_K^{\ast 2}\right)^{1/2} \ ,
\label{eq:omegaK}
\end{equation} 
where $X_0$ and $m_K^{\ast 2}$ are given by  Equations~(\ref{eq:x0}) and (\ref{eq:ekm2}), respectively.
The $\omega_K(\rho_B)$ decreases with an increase in $\rho_{\rm B}$ due to the $K$--$B$ scalar and vector attraction, while the kaon chemical potential $\mu_K$, which is equal to the charge chemical potential $\mu$ in $\beta$-equilibrated matter [Equation~(\ref{eq:chem})], increases with density. 
At a certain density,  $\omega_K(\rho_B)$ intersects with  $\mu$, where the condensed kaons spontaneously appear in the ground state through the weak reaction processes, $n+N\rightarrow p+N+K^-$, $l \rightarrow K^- +\nu_l $ ($l = e^-, \mu^-$), and strong reaction processes, 
$\Lambda\rightarrow p+K^-$, $\Xi^- \rightarrow \Lambda + K^-$, $\cdots$,  in the presence of hyperons. 
{Thus,} the onset density $\rho_B^c (K^-)$ for the $s$-wave kaon condensation is given by~\cite{mt92}:
\begin{equation}
\omega_K (\rho_B^c(K^-))=\mu 
\label{eq:onsetk}
\end{equation}
as a continuous phase transition.
The relaxation processes toward the equilibrated matter with KC are governed by the weak processes~\cite{mti2000,mti2000-2}.  

\begin{figure}[h]
\begin{minipage}[l]{0.50\textwidth}
\begin{center}
\includegraphics[height=0.27\textheight]{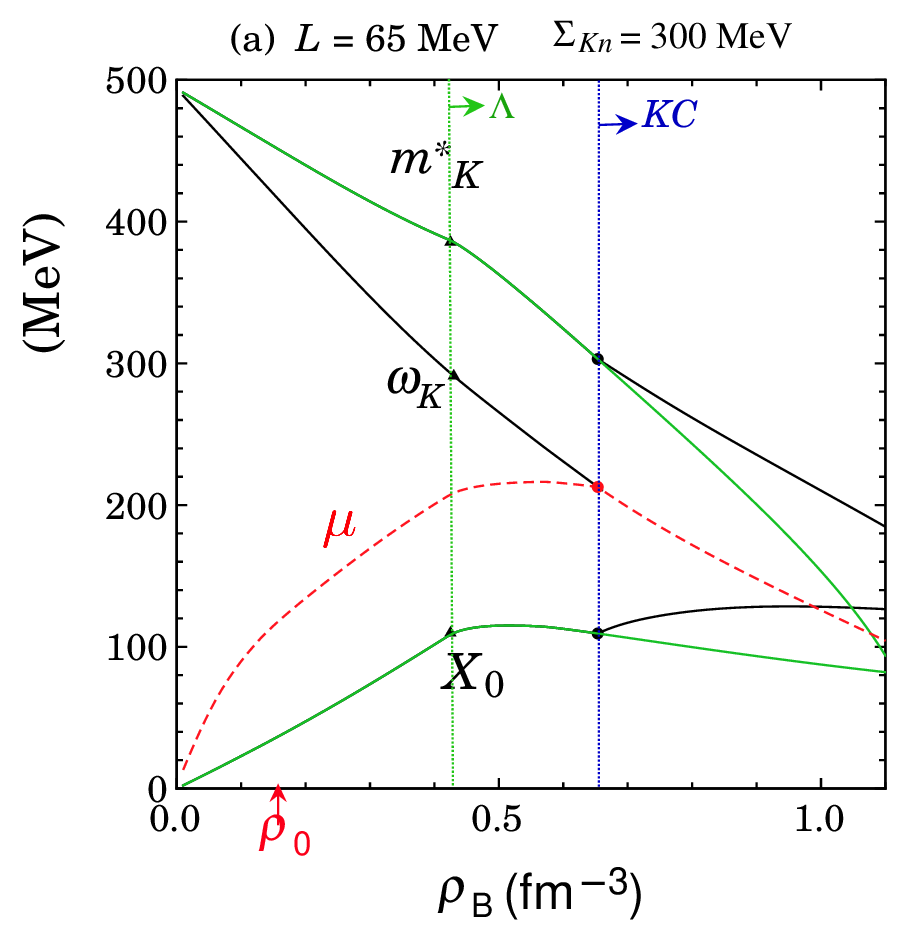}~
\end{center}
\end{minipage}~\hspace{0.5cm}
\begin{minipage}[r]{0.50\textwidth}
\begin{center}
\includegraphics[height=0.27\textheight]{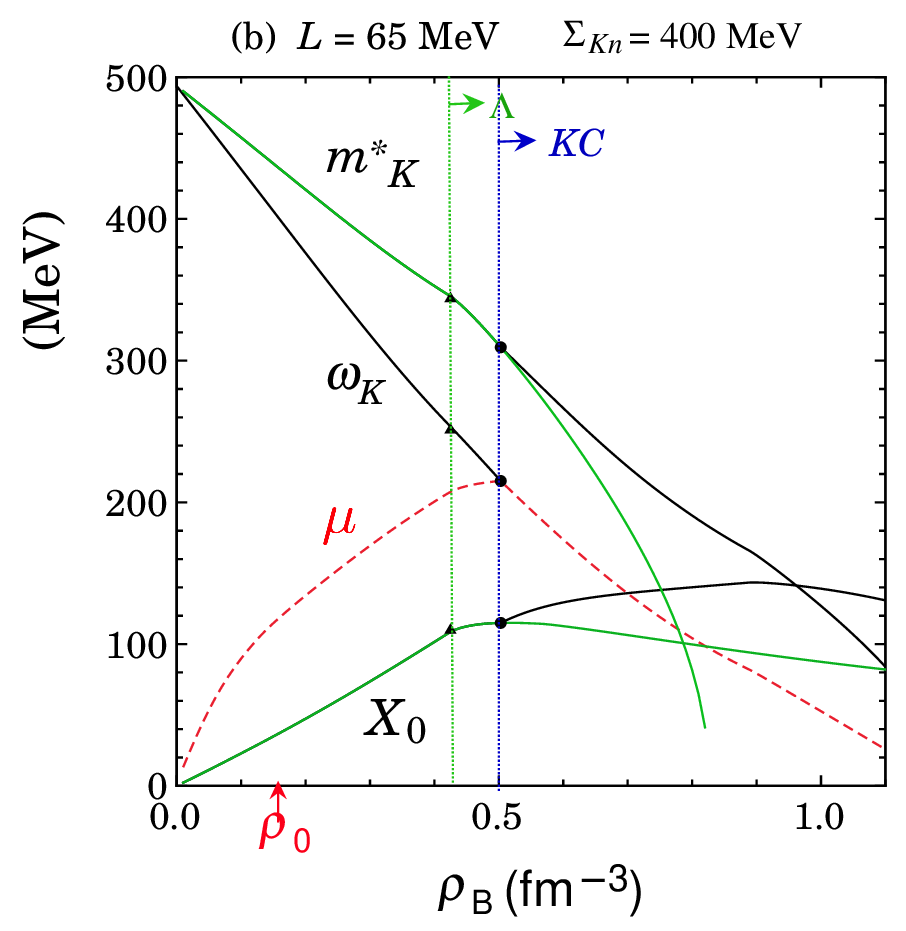}
\end{center}
\end{minipage}
\caption{(\textbf{a}) The lowest $K^-$ energy $\omega_K$, the effective mass of the $K^-$ meson $m_K^{\ast}$ defined by Equation~(\ref{eq:ekm2}), and the $X_0$ [Equation~(\ref{eq:x0})] as a function of the baryon number density $\rho_{\rm B}$ 
for $\Sigma_{Kn}$ = 300 MeV in the case of $L$ = 65 MeV. The $\rho_{\rm B}$-dependence of the charge chemical potential $\mu$ (=$\mu_e=\mu_\mu$ if muons are present) is also shown by the red dashed line. The filled triangle (filled circle) denotes the onset density of $\Lambda$ hyperon mixing, $\rho_{\rm B}^c(\Lambda)$ (the onset density of KC, $\rho_{\rm B}^c(K^-)$).  $\omega_K$ is equal to the charge chemical potential $\mu$ in the ($Y$+$K$) phase for $\rho_{\rm B}\geq\rho_{\rm B}^c(K^-)$. For a comparison, the density dependence of $m_K^{\ast}$ and $X_0$ in pure hyperon-mixed matter, where $\theta$ is set to be zero,  is also shown by the green lines.   (\textbf{b}) The same as (\textbf{a}), but for $\Sigma_{Kn}$ = 400 MeV. 
The filled triangle corresponds to the same onset density of $\Lambda$ as in (\textbf{a}). See the text for details. 
}
\label{fig:wk}
\end{figure}

 In Figure~\ref{fig:wk}, the lowest $K^-$ energy $\omega_K$ as a function of $\rho_{\rm B}$ is shown for (a) $\Sigma_{Kn}$ = 300~MeV and (b) $\Sigma_{Kn}$ = 400 MeV in the case of $L$ = 65 MeV. The charge chemical potential $\mu$ is also shown as a function of $\rho_{\rm B}$ by the red dashed line. The filled triangle denotes the onset density of $\Lambda$ hyperon mixing, $\rho_{\rm B}^c(\Lambda)$, at which hyperon ($\Lambda$) mixing starts in the normal neutron-star matter (nucleon matter). The filled circle denotes the $\rho_{\rm B}^c(K^-)$, at which KC is realized from hyperon ($\Lambda$ and/or $\Xi^-$)-mixed matter.~ 
The onset density of KC is read as  $ \rho_{\rm B}^c~(K^-)$ = 
(0.60--0.73)~fm$^{-3}$ [(3.7--4.6)~$\rho_0$] for $\Sigma_{Kn}$ = 300 MeV and 
 $\rho_{\rm B}^c~(K^-)$ = (0.49--0.52)~fm$^{-3}$ [(3.0--3.3)~$\rho_0$] for $\Sigma_{Kn}$ = 400 MeV, 
 within the range of the slope $L$ = (60--70) MeV. 
 For $\Sigma_{Kn}$ = 400 MeV,  $\omega_K$ is smaller at a given density than in the case of $\Sigma_{Kn}$ = 300 MeV due to the stronger $s$-wave $K$--$B$ scalar attraction, so that  $\rho_{\rm B}^c~(K^-)$ for $\Sigma_{Kn}$ = 400 MeV is lower than in the case of $\Sigma_{Kn}$ = 300 MeV. 
 In Table~\ref{tab:onset}, the onset densities $\rho_{\rm B}^c(\Lambda)$ and $\rho_{\rm B}^c(K^-)$ in the (ChL+MRMF+ UTBR+TNA) model for $\Sigma_{Kn}$ = 300 MeV and 400 MeV in the case of $L$ = (60, 65, 70) MeV are listed. For all of the cases of $L$ and $\Sigma_{Kn}$, the onset of $\Lambda$ mixing always precedes the onset of KC. 
\begin{table}[h]
\caption{The onset densities at which hyperon mixing starts and those of KC in the (ChL+MRMF+ UTBR+TNA) model for $\Sigma_{Kn}$ = 300 MeV and 400 MeV in the case of $L$=60, 65, and 70 MeV.  $\rho_{\rm B}^c(\Lambda)$ is the onset density of 
$\Lambda$ hyperons in the normal neutron-star matter, $\rho_{\rm B}^c(\Xi^- \ {\rm in} \ \Lambda )$ is one of the $\Xi^-$ hyperons in the $\Lambda$-mixed matter, $\rho_{\rm B}^c~(K^-)$ is one of the KC in the hyperon ($\Lambda$ and/or $\Xi^-$)-mixed matter, and $\rho_{\rm B}^c(\Xi^- \ {\rm in} \ K^- \Lambda )$ is one of the $\Xi^-$ hyperons in the KC phase in the $\Lambda$-mixed matter. }
\begin{center}
\begin{tabular}{ c | c  || c | c | c | c}
\hline
 $L$ & $\Sigma_{Kn}$  &  $\rho_{\rm B}^c(\Lambda)$ & $\rho_{\rm B}^c(\Xi^-~{\rm in}~\Lambda)$  &  $\rho_{\rm B}^c(K^-)$ & 
 $\rho_{\rm B}^c(\Xi^-~{\rm in}~{K^-\Lambda})$  \\
 (MeV) & (MeV)  &  (fm$^{-3}$) &  (fm$^{-3}$)   &  (fm$^{-3}$)   & (fm$^{-3}$)  \\ \hline\hline
60       & 300     &  0.466     &  $-$            & 0.598         & 1.04        \\
           & 400    &   0.466     &  $-$            & 0.486        &  0.994       \\ \hline
65       & 300     &  0.425     & 0.568         & 0.653         & $-$       \\
           & 400    &   0.425     & $-$             & 0.503         & 0.900         \\ \hline
70       & 300     &  0.397    &  0.516         & 0.733         & $-$          \\
           & 400    &   0.397    &  (0.516)         &  0.523        & 0.790          \\ \hline
\hline
\end{tabular}
\label{tab:onset}
\end{center}
\end{table}

In Ref.~\cite{mishra2010}, the modification of kaon properties in nucleonic and hyperon-mixed matter in neutron stars is investigated in the chiral SU(3) mean field model. The kaon self-energy includes the $K$--scalar-field interaction, $K$--vector-field interaction, and the range terms, corresponding to the case of our model interaction specified within chiral symmetry. The results on the density-dependence of the lowest kaon energy around and beyond the nuclear saturation density and the onset density of KC are also quantitatively similar to our case. On the other hand, 
the $K$--$K$ nonlinear self-interaction is naturally incorporated in our model as a consequence of the nonlinear representation of the $K$-field in the effective chiral Lagrangian. This nonlinear $K$--$K$ interaction may bring about any different aspect for the EOS beyond the onset density of KC.

\subsection{Interplay Between Kaons and Baryons Before and After the Onset of KC}
\label{subsec:interplay}

\begin{figure}[h]
\begin{minipage}[l]{0.50\textwidth}
\includegraphics[height=.31\textheight]{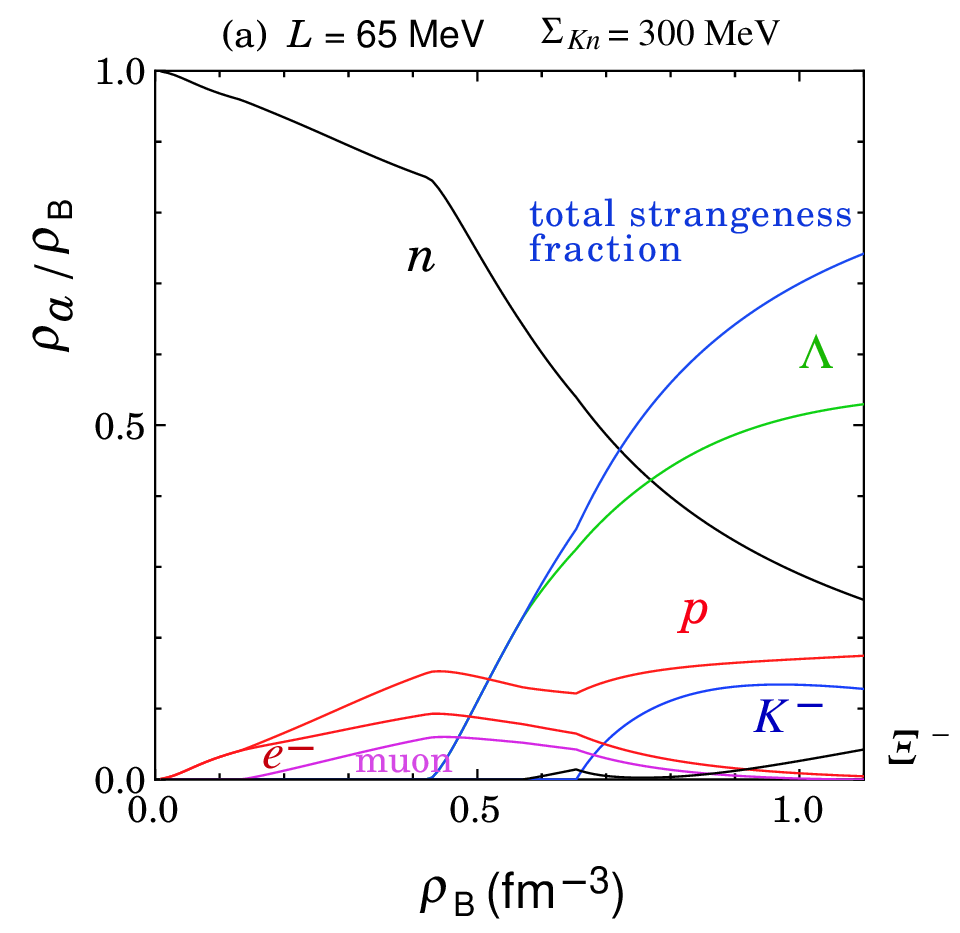}
\end{minipage}~
\begin{minipage}[r]{0.50\textwidth}
\includegraphics[height=.31\textheight]{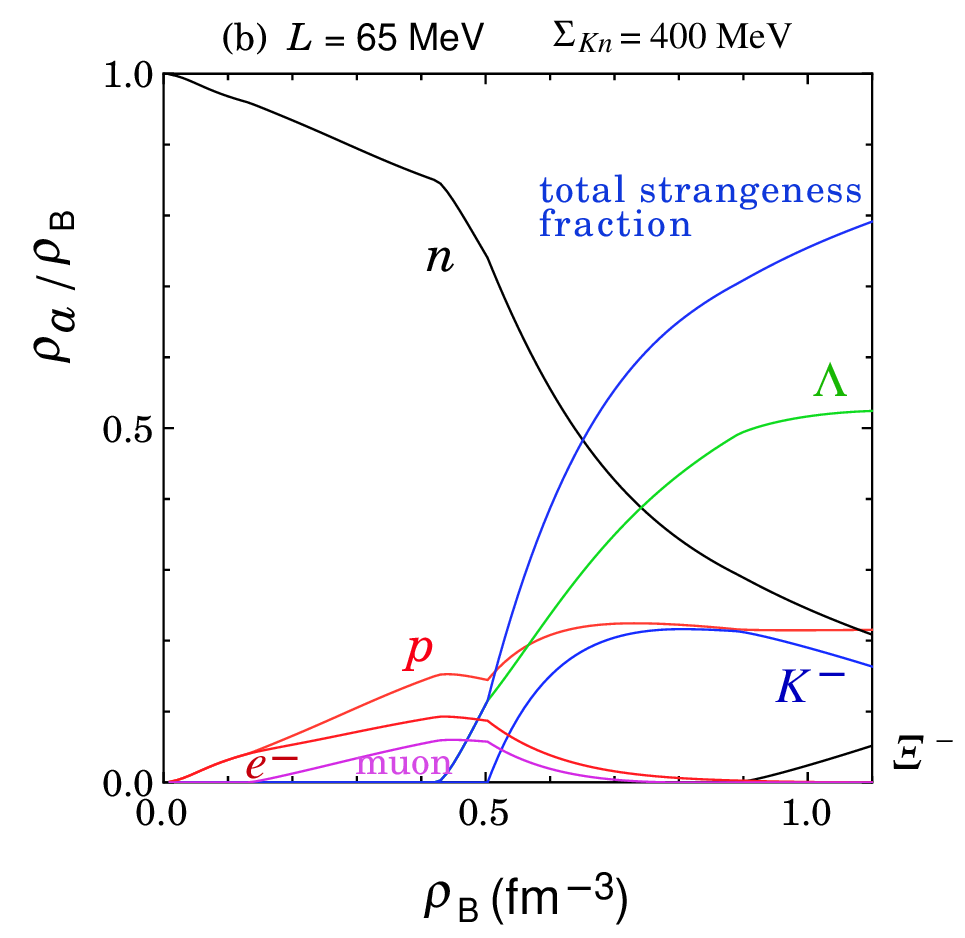}
\end{minipage}
\caption{(\textbf{a}) The particle fractions in the ($Y$+$K$) phase as functions of the baryon number density $\rho_{\rm B}$ for $\Sigma_{kn}$ = 300 MeV in the case of $L$ = 65 MeV. The total strangeness fraction is given by $(\rho_{K^-}+\rho_\Lambda+2\rho_{\Xi^-})/\rho_{\rm B}$. (\textbf{b}) The same as (\textbf{a}) but for  $\Sigma_{kn}$ = 400 MeV. } 
\label{fig:fractions}
\end{figure}

Together with  $\omega_K$, the density-dependence of  $X_0$ [Equation~(\ref{eq:x0})] and that of the ``effective mass'' $m_K^\ast$ of the $K^-$ meson [Equation~(\ref{eq:ekm2})] are shown in Figure~\ref{fig:wk}.
For reference, the particle fractions $\rho_a/\rho_{\rm B}$ before and after the onset of KC are shown as functions of $\rho_{\rm B}$ in Figure~\ref{fig:fractions}.
As seen in Figure~\ref{fig:fractions} and Table~\ref{tab:onset}, the $\Lambda$ mixing starts at a lower density than that of KC or $\Xi^-$ hyperons. Subsequently, the fraction of $\Lambda$ hyperons monotonically increases with density even after KC or $\Xi^-$ hyperons appear. 
Due to the appearance of $\Lambda$ hyperons, the nucleon (neutron and proton) fractions (and thus nucleon scalar densities) are suppressed as a result of baryon number conservation.
In pure hyperon-mixed matter, the decrease in the nucleon fractions leads to the suppression of the increase in $K$--$N$ attractive vector interaction simulated by $X_0$ [$\propto\left(\rho_p+\rho_n/2-\rho_{\Sigma^-}/2 - \rho_{\Xi^-}\right)/(2f^2)$ in Equation~(\ref{eq:x0})]. The appearance of the $\Xi^-$ hyperons also tends to work repulsively for the  $K$--$\Xi^-$ vector interaction, but the effect is negligible since $\rho_{\Xi^-}$ is tiny, even if the $\Xi^-$ hyperons are mixed in the range $\rho_{\rm B}^c(\Lambda)\lesssim \rho_{\rm B}\lesssim \rho_{\rm B}^c(K^-)$. On the other hand, $m_K^\ast$
reduces rapidly with density beyond the onset of $\Lambda$ mixing, due to the fact that the increase in $\Lambda$ scalar density overcomes the decrease in nucleon scalar densities and that the $K \Lambda$ sigma term is larger than the $KN$ ones.

Once KC appears, the KC and $\Xi^-$ hyperons compete against each other 
through the repulsive $K$--$\Xi^-$ vector interaction term in $X_0$, and the $\Xi^-$ fraction becomes small.   This competitive relation can also be seen from the interaction part of the $K^-$ number density $\rho_{K^-}$ [Equation~(\ref{eq:rhokc})]. As a result, the $X_0$ is slightly enhanced in comparison with the case of pure hyperon-mixed matter. On the other hand, the reduction in $m_K^\ast$ becomes moderate as compared with the case of pure hyperon-mixed matter, as seen in Figure~\ref{fig:wk}. Indeed, the $s$-wave $K$--$B$ scalar attraction is diminished according to the reduction in the total baryon scalar density in the presence of KC, since  a part of the strangeness is taken over by KC.

It is to be noted that the $\Sigma^-$ hyperons are not mixed over the relevant densities due to the strong repulsion of  $V_{\Sigma^-}^N$ in our model. 

The development of KC with an increase in the baryon density leads to an enhancement of the proton fraction
 so that the positive charge carried by protons compensates for the negative charge by KC, keeping charge neutrality. 
On the other hand, the lepton ($e^-$, $\mu^-$) fractions are suppressed after the appearance of KC as well as $\Lambda$ hyperons, since the negative charge carried by leptons is replaced by that of KC,  avoiding the cost of the degenerate energy of leptons. The ($Y$+$K$) phase becomes almost lepton-less at high densities (see Figure~\ref{fig:fractions}). 
As a consequence, the charge chemical potential $\mu$ [=$(3\pi^2\rho_e)^{1/3}$ ] decreases steadily as the density increases after the onset of KC and has a value with $\mu\lesssim O(m_\pi)$ (see Figure~\ref{fig:wk}). 
These features concerning proton and lepton fractions and the charge chemical potential are characteristic of the hadron phase in the presence of KC.
The total strangeness is carried mainly by $\Lambda$ hyperons and KC in the ($Y$+$K$) phase with a minor fraction of $\Xi^-$ hyperons at high densities.

\section{EOS and Structure of Neutron Stars with the (\boldmath{$Y$}+$K$) Phase}
\label{sec:MR}

In this section, we first summarize the results on the stiffness of the EOS with the ($Y$+$K$) phase. 
The energy contribution from the UTBR per baryon, $ {\cal E}~({\rm UTBR})/\rho_{\rm B}$, which is roughly proportional to ${\rho_{\rm B}}^2$, has a sizable contribution to the total energy and results in the stiffening of the EOS at high densities. 
The $E$~(two-body) [$={\cal E}_{B,M}/\rho_{\rm B}$] also brings about repulsive energy as large as $E$~(UTBR) until the onset of KC. 
Beyond the onset density of KC, the $E$~(two-body) turns to decrease with density due to the attraction from the $s$-wave $K$--$B$ interaction for both cases of $\Sigma_{Kn}$, until it increases again at higher density.  
On the other hand, the $E$~(KC) [$={\cal E}_K/\rho_{\rm B}$], composed of kinetic and mass terms of KC, increases with baryon density. 
The sum of $E$~(two-body) and $E$~(KC) results in positive energy, which increases with baryon density and works to stiffen the EOS as much as $E$~(UTBR). 
See Ref.~\cite{muto2024} for more details.

Second, we discuss the effects of KC on the static properties of compact stars such as the gravitational mass $M$--radius $R$ relations. They are obtained by solving the Tolman--Oppenheimer--Volkoff equation with the EOS including the ($Y$+$K$) phase. For the low-density region $\rho_{\rm B} < $ 0.10~fm$^{-3}$, which is below the density of uniform matter, we utilize the EOS of Ref.~\cite{BPS1971} and combine it with the EOS obtained in our model for $\rho_{\rm B}\geq$ 0.10~fm$^{-3}$. 

 In Table~\ref{tab:MR}, some critical gravitational masses and their radii are summarized for $\Sigma_{Kn}$ = 300 MeV and 400 MeV in the case of $L$ = (60, 65, 70) MeV. 
$M^c(\Lambda)$ and $R^c(\Lambda)$ [$M^c(K^-)$ and $R^c(K^-)$] are the mass and radius of the neutron star, where the central density attains the onset density of the $\Lambda$-hyperons, $\rho_{\rm B}^c(\Lambda)$ [the onset density of KC, $\rho_{\rm B}^c(K^-)$]. $M_{\rm max}$ and $R(M_{\rm max})$ are the maximum mass of the neutron star and its radius. 
\begin{table}[h]
\caption{Some critical gravitational masses in the unit of the solar mass $M_\odot$ and their radii of neutron stars for $\Sigma_{Kn}$ = 300 MeV and 400 MeV in the case of $L$ = (60, 65, 70) MeV, obtained with the (ChL+MRMF+UTBR+TNA) model. The $M^c(\Lambda)$ and $R^c(\Lambda)$ [$M^c(K^-)$ and $R^c(K^-)$] are the mass and radius of the neutron star, where the central density reaches the onset density of the $\Lambda$-hyperons, $\rho_{\rm B}^c(\Lambda)$ [the onset density of KC, $\rho_{\rm B}^c(K^-)$]. 
$M_{\rm max}$ and $R(M_{\rm max})$ are the maximum mass of the neutron star and its radius. \\ }
\begin{tabular}{  c | c || c | c || c | c || c | c }
\hline
 $L$ & $\Sigma_{Kn}$ & $M^c(\Lambda)/M_\odot $ & $R^c(\Lambda)$ & $M^c(K^-)/M_\odot $ & $R^c(K^-)$ & $M_{\rm max}/M_\odot $ & $R(M_{\rm max})$  \\
 (MeV) & (MeV)          &                                           & (km)                   &                               & (km)           
&                         & (km)                      \\
\hline\hline
 60 & \begin{tabular}{c} 
 300   \\
 400   \\
\end{tabular}             & 1.448                                    &  12.33                & 
\begin{tabular}{c} 
1.742     \\
1.452     \\
\end{tabular}  
& \begin{tabular}{c}
12.11     \\
12.33     \\
\end{tabular}
&  \begin{tabular}{c}
2.035       \\
1.993       \\
\end{tabular}
& \begin{tabular}{c} 
10.02    \\
 9.48  \\
\end{tabular}  \\\hline\hline
 65 & \begin{tabular}{c}
 300     \\
 400    \\
 \end{tabular}
& 1.508                                    &  12.78                
& \begin{tabular}{c}
1.961      \\
1.737      \\
\end{tabular}    
& \begin{tabular}{c}
12.29     \\
12.68     \\
\end{tabular}
& \begin{tabular}{c}
2.124      \\ 
2.076     \\
\end{tabular}
& \begin{tabular}{c}
10.76  \\
10.29  \\
\end{tabular} \\\hline\hline      
 70 & \begin{tabular}{c}
 300     \\
 400     \\
 \end{tabular}
& 1.582      &  13.15              
& \begin{tabular}{c}
2.139      \\
1.915  \\
\end{tabular}
&   \begin{tabular}{c}
12.24      \\ 
12.97    \\
\end{tabular} 
&\begin{tabular}{c}
 2.200        \\
2.155         \\
\end{tabular}
& \begin{tabular}{c}
11.31       \\
 11.06      \\
\end{tabular} \\ \hline
 \end{tabular}
\label{tab:MR}
\end{table}
The mass of the neutron star where the central density reaches the onset density $\rho_{\rm B}^c(K^-)$ is (1.74--2.14)$M_\odot$ for $L$ = (60--70) MeV in the case of $\Sigma_{Kn}$ = 300 MeV, 
and (1.45--1.92)$M_\odot$ for $L$ = (60--70) MeV in the case of $\Sigma_{Kn}$ = 400 MeV.  

Observationally, neutron stars as large as 2~$M_\odot$ ($M_\odot$ being the solar mass) have been detected~\cite{demo10,fonseca2016,ant13,c2020,fonseca2021}. Both the mass and radius have been detected from X-ray observation by the Neutron star Interior Composition ExploreR (NICER): for the pulsar PSR~J0740+6620 with $M_{\rm obs.}$ = 2.08$M_\odot$, $R_{\rm obs.}$ = (12.35 $\pm$ 0.75)~km~\cite{miller2021} and $M_{\rm obs.}$ = (2.072$^{+0.067}_{-0.066}$)~$M_\odot$, $R_{\rm obs.}$ = (12.39$^{+1.30}_{-0.98}$)~km~\cite{riley2021}, and for PSR~J0030+0451 with $M_{\rm obs.}$ = (1.34$^{+0.15}_{-0.16}$)$M_\odot$, $R_{\rm obs.}$ = (12.71$^{+1.14}_{-1.19}$)~km~\cite{riley2019}, and $M_{\rm obs.}$ = (1.44$^{+0.15}_{-0.14}$)$M_\odot$, $R_{\rm obs.}$ = (13.02$^{+1.24}_{-1.06}$)~km~\cite{miller2019}. 
The curves of $M$--$R$ relations based on our EOS in the case of $L$ = (65, 70) MeV pass through the above constrained regions. In particular, the maximum masses with the ($Y$+$K$) phase in the core are consistent with recent observations of massive neutron stars in both the cases of \mbox{$\Sigma_{Kn}$ = 300 MeV} and 400 MeV for $L$ =(65, 70) MeV. However, the masses within the causal limit for $\Sigma_{Kn}$ = 400 MeV and $L$ = 60 MeV do not reach the range allowable from the observations of most massive neutron stars. 
In our model, the larger values of the slope $L \gtrsim$ 60~MeV are preferred in order to obtain observed massive neutron stars~\cite{muto2024}.

\section{Quark Condensates in the (\boldmath{$Y$+$K$}) Phase and Relevance to Chiral Restoration}
\label{sec:qcondensates}

Following the preceding results on the properties of the ($Y$+$K$) phase, we discuss the effects of the ($Y$+$K$) phase on chiral restoration in dense matter by obtaining the quark condensates in the ($Y$+$K$) phase within the mean field approximation. 
The quark condensate in KC  (for $q=u, d, s$) is {expressed} as: 
\begin{eqnarray}
\langle \bar q q \rangle_{\rm KC}&\equiv& \langle {\rm KC}|\bar q q|{\rm KC}\rangle \cr
&=& \langle {\rm KC}|d\hat {\cal H}/dm_q |{\rm KC}\rangle  \ ,
\label{eq:qcondensates}
\end{eqnarray}
where $|{\rm KC} \rangle$ is the kaon-condensed eigenstate of the total Hamiltonian 
$\hat {\cal H}$ ($\displaystyle =\sum_{q=u,d,s} m_q\bar q q + \cdots$) with the eigenvalue of the ground state energy density ${\cal E}_{\rm g.r.}$, ~i.e.,~${\cal \hat H}|{\rm KC}\rangle = {\cal E}_{\rm g.r.}|{\rm KC}\rangle$, and $\langle {\rm KC}|{\rm KC}\rangle=1$.
With the Feynman--Hellmann theorem~\cite{cohen1992}, one obtains
\begin{eqnarray}
\langle\bar q q \rangle_{\rm KC}&=&d\langle{\rm KC}|\hat {\cal H}|{\rm KC}\rangle/dm_q \cr
&=&d{\cal E}_{\rm g.r.}/dm_q  \ .
\label{eq:FHKC}
\end{eqnarray}
Throughout this paper, ${\cal E}_{\rm g.r.}$ is approximated to ${\cal E}_{\rm g.r.}\simeq{\cal E}_0 + \Delta {\cal E}_0+{\cal E}$, where ${\cal E}_0$ is the vacuum energy, $\Delta {\cal E}_0$ is the energy shift due to vacuum polarization in the presence of baryonic matter, and ${\cal E}$ [Equation~(\ref{eq:total-edensity})] is the energy density of the ($Y$+$K$) phase in the mean field approximation. In the following, we further neglect  $\Delta {\cal E}_0$. Then, the quark condensate associated with the charged kaon condensation reads as follows:
\begin{eqnarray}
\langle\bar u u +\bar s s\rangle_{\rm KC}&=&\sum_{q=u, s}d({\cal E}_0+{\cal E})/dm_q \cr
&=& \langle\bar u u +\bar s s\rangle_0+ {\sum_{q=u, s} d{\cal E}/dm_q  \ ,}
\label{eq:qcKCE}
\end{eqnarray}
where the first term in the second line on the r.~h.~s. is the quark condensate in the vacuum,
and 
the second term is the contribution from the total energy density ${\cal E}$ [Equation~(\ref{eq:total-edensity})]. The latter is further separated into the one from the kaon mass term [Equation~(\ref{eq:ekfinal})] and 
the one from the baryonic energy Equation~(\ref{eq:ebm}): $\displaystyle \sum_{q=u, s} d{\cal E}/dm_q = \sum_{q=u, s} d{\cal E}_K/dm_q + \sum_{q=u, s} d{\cal E}_{B, M} / dm_q$. 
With the use of the kaon rest mass [Equation~(\ref{eq:kmass})], one obtains the following:
\begin{equation}
\sum_{q=u, s} d{\cal E}_K/dm_q =  f^2(1-\cos\theta) \sum_{q=u, s} dm_K^2/dm_q = \frac{2f^2m_K^2}{m_u+m_s}(1-\cos\theta) \ .
\label{eq:qcK}
\end{equation}
 With the use of Equation~(\ref{eq:ebm}), the contribution from ${\cal E}_{B, M}$ is given as:
\begin{equation}
\sum_{q=u, s} d{\cal E}_{B, M} / dm_q = \sum_{q=u, s} \sum_b\rho_b^s \left(\partial\widetilde M_b^\ast / \partial m_q\right)+\sum_{q=u, s}\left(m_\sigma^2 \sigma \partial\sigma / \partial m_q+m_{\sigma^\ast}^2 \sigma^\ast \partial\sigma^\ast / \partial m_q \right) \ , 
\label{eq:qcBM}
\end{equation}
where the factor $\partial\widetilde M_b^\ast / \partial m_q$ is written with the use of Equation~(\ref{eq:wtmb}) as:
\begin{equation}
\partial\widetilde M_b^\ast / \partial m_q = \langle \bar q q\rangle_b -\left(\partial\Sigma_{Kb} / \partial m_q \right) (1-\cos\theta) - \left(g_{\sigma b}\partial\sigma / \partial m_q
+g_{\sigma^\ast b}\partial\sigma^\ast / \partial m_q\right) \ .
\label{eq:qcSigma}
\end{equation}
Substitution of Equation~(\ref{eq:qcSigma}) into Equation~(\ref{eq:qcBM}) leads to
\begin{equation}
\sum_{q=u, s} d{\cal E}_{B, M} / dm_q = \sum_{q=u, s} \sum_b\rho_b^s \Big\lbrace \langle \bar q q\rangle_b -\left( \partial\Sigma_{Kb} / \partial m_q \right) (1-\cos\theta) \Big\rbrace \ , 
\label{eq:qcBM2}
\end{equation}
where each term proportional to $\partial\sigma / \partial m_q$ and $\partial\sigma^\ast / \partial m_q$ is shown to vanish separately with the help of the equations of motion for the $\sigma$ mean field [Equation~(\ref{eq:cieom1})] and $\sigma^\ast$ mean field [Equation~(\ref{eq:cieom2})], respectively. 
Noting that 
\begin{equation}
 \sum_{q=u, s} \langle \bar q q\rangle_b=  \sum_{q=u, s} \partial\Sigma_{Kb} / \partial m_q =\frac{2\Sigma_{Kb}}{m_u+m_s} \ ,
 \label{eq:qcqq}
 \end{equation}
one can write Equation~(\ref{eq:qcBM2}) simply as:
\begin{equation}
\sum_{q=u, s} d{\cal E}_{B, M} / dm_q = \frac{2\cos\theta}{m_u+m_s}\sum_b\rho_b^s\Sigma_{Kb} \ .
\label{eq:qcBM3}
\end{equation}
From Equations~(\ref{eq:qcK}) and (\ref{eq:qcBM3}), one obtains the following:
\begin{equation}
\sum_{q=u, s} d{\cal E}/dm_q = \frac{2f^2m_K^2}{m_u+m_s}\Bigg\lbrace 1-\left(\frac{m_K^{\ast 2}}{m_K^2}\right)\cos\theta \Bigg\rbrace 
\label{eq:qcE}
\end{equation}
with the use of Equation~(\ref{eq:ekm2}) for $m_K^{\ast 2}$. 

\begin{figure}[h]
\begin{minipage}[l]{0.50\textwidth}
\includegraphics[height=.32\textheight]{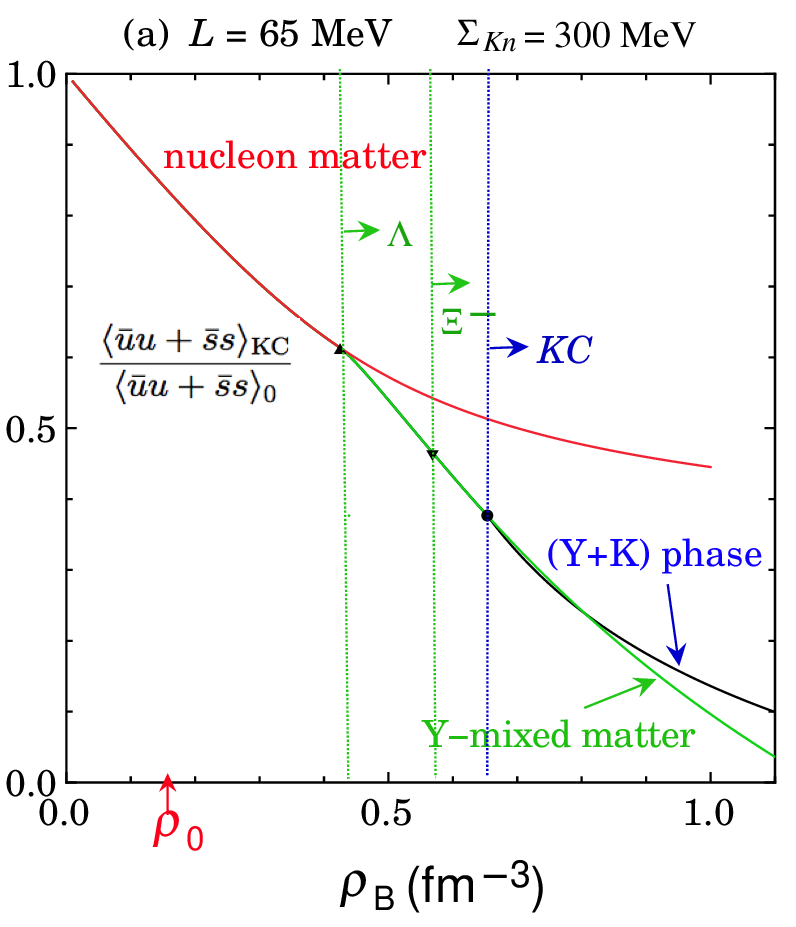}
\end{minipage}~
\begin{minipage}[r]{0.50\textwidth}
\includegraphics[height=.32\textheight]{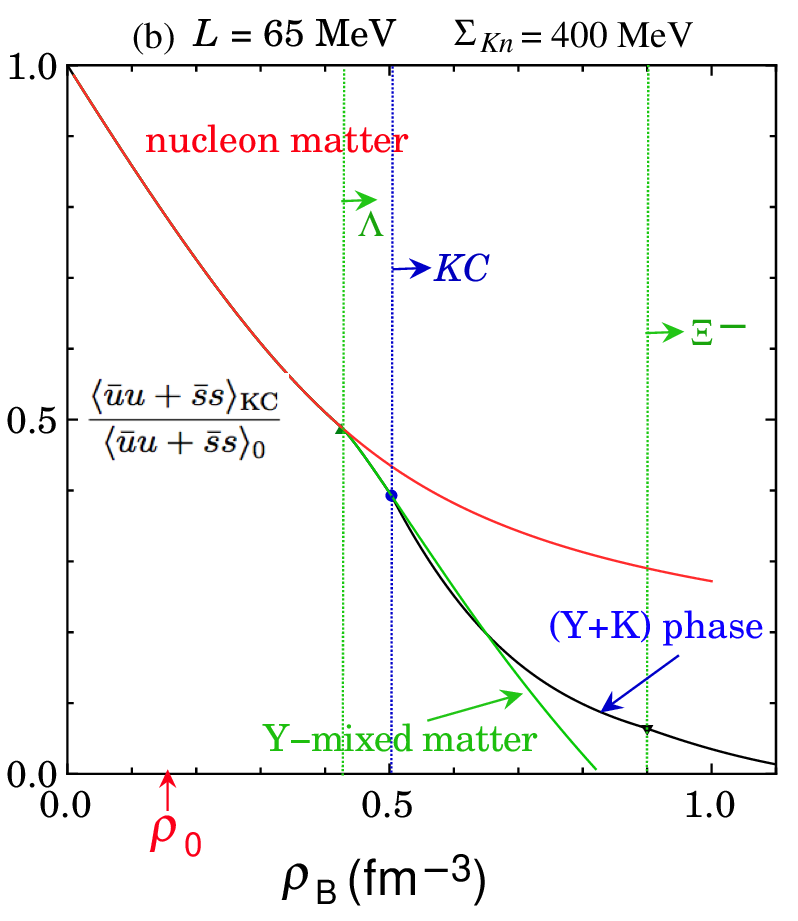}
\end{minipage}
\caption{(\textbf{a}) The ratio of the quark condensate in the ($Y$+$K$) phase to the vacuum quark condensate, $\langle\bar u u +\bar s s\rangle_{\rm KC}/\langle\bar u u +\bar s s\rangle_0$, as a function of $\rho_{\rm B}$ for $L$ = 65 MeV and $\Sigma_{Kn}$ = 300 MeV is shown by the black line. For comparison, the one in pure hyperon-mixed matter without KC by setting $\theta = 0$ is shown by the green line and the one for nucleon matter, i.e.,~pure neutron-star matter without hyperon mixing and KC, is shown by the red line. (\textbf{b}) The same as (\textbf{a}) but for $\Sigma_{Kn}$ = 400 MeV. \\}
\label{fig:qckc}
\end{figure}

The vacuum condensate $\langle\bar u u +\bar s s\rangle_0$ is related to {the meson decay constant $f$ by} the Gell-Mann--Oakes--Renner (GOR) relation:
\begin{equation}
\langle\bar u u +\bar s s\rangle_0=-\frac{2f^2 m_K^2}{m_u+m_s} \ . 
\label{eq:GOR}
\end{equation}
Substituting Equations~(\ref{eq:qcE}) and (\ref{eq:GOR}) into Equation~(\ref{eq:qcKCE}), one finally obtains
\begin{equation}
\frac{\langle\bar u u +\bar s s\rangle_{\rm KC}}{\langle\bar u u +\bar s s\rangle_0}=\left(\frac{m_K^{\ast 2}}{m_K^2}\right)\cos\theta  \ .
\label{eq:qckcratio}
\end{equation}
Thus, the density-dependence of the quark condensate is determined by the $s$-wave $K$--$B$ scalar interaction simulated by the $Kb$ sigma terms $\Sigma_{Kb}$ within the mean field approximation. 
In Figure~\ref{fig:qckc}, the ratio of the quark condensate in the ($Y$+$K$) phase to the vacuum quark condensate, $\langle\bar u u +\bar s s\rangle_{\rm KC}/\langle\bar u u +\bar s s\rangle_0$, is shown as a function of the baryon number density $\rho_{\rm B}$ for $L$ = 65 MeV with $\Sigma_{Kn}$ = 300 MeV for (a) and $\Sigma_{Kn}$ = 400 MeV for (b). 
One can see that the appearance of hyperons leads to a larger decrease in the condensate  
as density increases in comparison with the case of non-strangeness matter (nucleon matter). In the presence of KC, the decrease in the quark condensate is also enhanced in comparison with the case of nucleon matter by the reduction factor $\cos\theta$ in Equation~(\ref{eq:qckcratio}), while the decrease is moderated in comparison with the case of pure hyperon-mixed matter, since the $s$-wave $K$--$B$ scalar attractive interaction is weakened as a result of the competing effect between hyperons and KC. The decrease in the quark condensate is {larger} for $\Sigma_{Kn}$ = 400 MeV than for $\Sigma_{Kn}$ = 300 MeV. 
{Thus,} it may be concluded that the
appearance of strangeness in the form of hyperon mixing and KC in dense matter assists in the restoration of 
chiral symmetry. 

It is to be noted that a contribution from particle--hole correlations by baryons and mesons beyond the mean field approximation is not taken into account in the present form of the $\bar q q$ condensates, Equation~(\ref{eq:qckcratio}), nor the vacuum polarization effect on the $\bar q q$ condensates, $d(\Delta {\cal E}_0)/dm_q$, through the modification of the Dirac sea in the presence of the Fermi sea. These effects should be taken into account for future study.

\section{Summary and Outlook}
\label{sec:summary}

We have overviewed the properties of the coexistent phase of kaon condensates and hyperons [($Y$+$K$) phase] by the use of the interaction model based on the effective chiral Lagrangian (ChL) for $K$--$B$ and $K$--$K$ interactions combined with the minimal relativistic mean field theory (MRMF) for two-body baryon interaction, taking into account the universal three-baryon repulsion (UTBR) and the phenomenological three-nucleon attraction (TNA), referring to the results in Ref.~\cite{mmt2021,muto2024}. 
 The interplay between KC and hyperons and the resulting onset mechanisms of KC in hyperon-mixed matter and the EOS with the ($Y$+$K$) phase have been clarified within the (ChL+MRMF+UTBR+TNA) model. 
The EOS and the resulting mass and radius of compact stars within a hadronic picture accompanying the ($Y$+$K$) phase are consistent with recent observations of massive neutron stars. 

We have figured out the close relations between the $s$-wave KC in the ($Y$+$K$) phase and the quark ($\bar q q$) condensates in the context of chiral symmetry and its spontaneous and explicit breaking.
One is the estimation of the quark {contents} inside the baryon, which is connected to the $Kb$ sigma term as one of the driving forces for the $s$-wave KC. By taking into account the nonlinear effect with respect to the strangeness quark mass beyond the chiral perturbation, we have obtained the allowable range of the $Kn$ sigma term, for a given $\pi N$ sigma term, which is suggested from phenomenological analyses of the $\pi$-$N$ scattering experiments, and the small $\bar s s$ strangeness {content} in the nucleon, which is suggested from the recent lattice QCD results. As a result, the values of  $\Sigma_{Kn}$ = (300--400) MeV have been adopted as reasonable values in the paper. 

Second, we have obtained the $\bar q q$ condensates in the ($Y$+$K$) phase in the mean field approximation. It has been shown that 
 both the appearance of strangeness in the form of hyperon mixing and KC in dense matter assists the restoration of chiral symmetry. 

As an outlook with regard to the realistic EOS including various aspects of MC over the whole densities, it is suggested from heavy-ion collision experiments that the EOS in the SNM or in pure neutron matter may be softer for $\rho_{\rm B}$~(2--4.5)~$\rho_0$~\cite{dll2002}. Pion condensation (PC), which may be realized at rather low densities $\rho_{\rm B}\gtrsim 2\rho_0$, may have a role as a softening mechanisms of the EOS at the relevant densities~\cite{mtt93}.
A possible coexistence of PC and KC ($\pi$-$K$ condensation) may be a realistic form of hadronic phase for keeping from the assumption of the UTBR. 
In the ground state of the $\pi$-$K$ condensed phase in neutron-star matter, the energy eigenstates for baryons are given by quasi-baryonic states with the superposition of neutron and proton states under the $p$-wave {charged pion} condensates. In such a case, the ground state is occupied solely by the lower-energy eigenstates of the quasi-baryons as a result of the level repulsion, forming the one-Fermi sea, which may help resolve the assumption of the universal strengths between different species of baryons for the UTBR. 

 On the other hand, there have been extensive studies on MC in quark matter~\cite{mannarelli2019}: PC in the Nambu--Jona--Lasinio (NJL) model~\cite{a2009}, PC in the NJL model with chiral imbalance~\cite{kkz2019,kkz2020}, PC and KC in chiral perturbation theory~\cite{aa2020,aa2020-2}, KC in the Ginzburg--Landau model with axial anomaly~\cite{sst2011}, etc.
 Recently, hadron--quark crossover has been proposed to obtain massive compact stars compatible with observations~\cite{mht2013,mht2016,baym2019,kojo2021,fujimoto2022}. In this context, the connection of hadronic matter including the ($Y$+$K$) phase to quark matter at high densities may be possible. 
 Specifically, there may be similarity and difference between the alternating layer spin (ALS) structure accompanying $\pi^0$ condensation in hadronic matter~\cite{kmttt1993} and a dual chiral density wave (DCDW) in quark matter~\cite{tn2004,nt2005}. It is also an open problem how the ($Y$+$K$) phase is connected to KC in the color--flavor-locked (CFL) phase~\cite{son2000,bedaque2002,kr2002}. 
 Toward a unified understanding of meson condensation in both the hadronic phase and quark phase, 
 tje correspondence between chiral dynamics in both phases should be clarified on the assumption that there remain various hadronic excitation modes even in quark matter. For example, the following issues may be left as a future elucidation whether there are relevant meson--quark interactions for MC in quark matter, corresponding to the $s$-wave $K$--$B$ scalar and vector interactions as the driving force of the $s$-wave KC, and those corresponding to the $p$-wave $\pi N$ interaction as the driving force of the $p$-wave PC in hadronic matter. 
Multi-quark interaction may be responsible for stiffening the EOS of MC in quark matter. Such a repulsion might correspond to the UTBR, which is introduced phenomenologically in order to solve the significant  softening problem stemming from the appearance of KC and hyperon mixing in hadronic matter.

\vspace{+6pt}

\acknowledgments{The authors thank H.~Sotani, N.~Yasutake, and A.~Dohi for their useful comments and interest in this work. The work is financially supported by the Chiba Institute of Technology. }


\begin{thebibliography}{999}

\bibitem{sawyer1972} Sawyer, R.F. Condensed $\pi^-$ phase in neutron-star matter.
\emph{Phys.~Rev.~Lett.}~{\bf 1972}, \emph{29}, 382.  

\bibitem{scalapino1972} Scalapino, D.J. $\pi^-$ condensate in dense nuclear matter. \emph{Phys.~Rev.~Lett.}~{\bf 1972}, \emph{29},~392. 

\bibitem{migdal1978} Migdal, A.B. Pion fields in nuclear matter. \emph{Rev.~Mod.~Phys.}~{\bf 1978}, \emph{50},~107. 

\bibitem{migdal1990} Migdal, A.B.; Saperstein, E.E.; Troitsky, M.A.; Voskresensky, D.N. Pion degrees of freedom in nuclear matter. \emph{Phys.~Rep.}~{\bf 1990}, \emph{192},~179. 

\bibitem{baym1979} Baym, G.; Campbell, D.K. {\it Mesons and Nuclei}; Rho, M., Wilkinson, D.H., Eds.; North Holland, Amsterdam, 1979; Volume~III, p.~1031.

\bibitem{kmttt1993} Kunihiro, T.; Muto, T.; Takatsuka, T.; Tamagaki, R.; Tatsumi, T. 
Various phases in high-density nuclear matter and neutron stars. \emph{Prog.~Theor.~Phys.~Suppl.}~{\bf {1993}}, \emph{112}, 1.

\bibitem{kn86} Kaplan, D.B.; Nelson, A.E. Strange goings on in dense nucleonic matter. 
\emph{Phys.~Lett. B}~{\bf 1986}, \emph{175},~57. 

\bibitem{t88} Tatsumi, T. K-on condensation and cooling of neutron stars. \emph{Prog.~Theor.~Phys.}~{\bf 1988}, \emph{80},~22.

\bibitem{mt92} Muto, T.; Tatsumi, T. Theoretical aspects of kaon condensation in neutron matter. \emph{Phys.~Lett. B}~{\bf 1992}, \emph{283},~165.

\bibitem{m93} Muto, T. Role of weak interaction on kaon condensation in neutron matter -- A result with hyperon excitations. \emph{Prog.~Theor.~Phys.}~{\bf 1993}, \emph{89},~415.

\bibitem{mtt93}
Muto, T.; Tamagaki, R.; Tatsumi, T. A chiral symmetry approach to meson condensations. \emph{Prog.~Theor.~Phys.~Suppl.} {\bf 1993}, \emph{112},~159. 

\bibitem{mttt1993}
Muto, T.; Takatsuka, T.; Tamagaki, R.; Tatsumi, T. Implications of various hadron phases to neutron star phenomena. \emph{Prog.~Theor.~Phys.~Suppl.} {\bf 1993}, \emph{112},~221.

\bibitem{tpl94} Thorsson, V.; Prakash, M.; Lattimer, J.M. Composition, structure and evolution of neutron stars with kaon condensates.
\emph{Nucl.~Phys. A}~{\bf 1994}, \emph{572},~693.

\bibitem{kvk95} Kolomeitsev, E.E.; Voskresensky, D.N.; K{\"a}mpfer, B. Kaon polarization in nuclear matter. \emph{Nucl.~Phys.~A} {\bf 1995}, \emph{588}~889. 

\bibitem{lbm95} Lee, C.-H.; Brown, G.E.; Min, D.-P.; Rho, M. An Effective chiral Lagrangian approach to kaon - nuclear interactions: Kaonic atom and kaon condensation. \emph{Nucl.~Phys.~A}{\bf  1995}, \emph{585},~401. 

\bibitem{lee1996} Lee, C.-H. Kaon condensation in dense stellar matter. \emph{Phys.~Rep.} {\bf 1996}, \emph{275}, 255.

\bibitem{prakash1997} Prakash, M.; Bombaci, I.; Prakash, M.; Ellis, P.J.; Lattimer, J.M.; Knorren. Composition and Structure of Protoneutron Stars.\emph{Phys.~Rep.}~{\bf 1997}, \emph{280}, 1.

\bibitem{tstw98} Tsushima, K.; Saito, K.; Thomas, A.W.; Wright, S.V. In-medium kaon and antikaon properties in the quark-meson coupling model. 
\emph{Phys.~Lett.~B} {\bf 1998}, \emph{429},~239.

\bibitem{fmmt1996} Fujii, H.; Maruyama, T.; Muto, T.; Tatsumi, T. Equation of state with kaon condensates and neutron stars. \emph{Nucl.~Phys.~A} {\bf 1996}, \emph{ 597},~645.

\bibitem{g2001} {Glendenning, N.K.; Schaffner-Bielich, J. First order kaon condensate. \emph{Phys.~Rev.~C} {\bf 1999}, \emph{60},~025803.}

\bibitem{maxwell1977} {Maxwell, O.V.; Brown, G. E.; Campbell, D. K.; Dashen, R. F.; Manassah, J. T. 
Beta decay of pion condensates as a cooling mechanism for neutron stars. 
 \emph{Astrophys.~J.}~{\bf 1977}, \emph{216},~77.}

\bibitem{bkpp1988} {Brown, G.E.; Kubodera, K.; Page, D.; Pizzecherro, P. Strangeness condensation and cooling of neutron stars. 
\emph{Phys.~Rev.~D} {\bf 1988}, \emph{37}, 2042. }

\bibitem{fmtt1994} {Fujii, H.; Muto, T.; Tatsumi, T.; Tamagaki, R. Effects of weak interaction on kaon condensation and cooling of neutron stars. \emph{Nucl.~Phys.~A} {\bf 1994}, \emph{571},~758.
 
 \bibitem{fmtt1994-2} Fujii, H.; Muto, T.; Tatsumi, T.; Tamagaki, R. Effects of symmetry energy on the direct URCA process in the kaon condensed phase.
\emph{Phys.~Rev.~C} {\bf 1994}, \emph{50},~3140.
}
\bibitem{ekp95} Ellis, P.J.; Knorren, R.; Prakash, M. Kaon condensation in neutron star matter with hyperons. 
\emph{Phys.~Lett.~B} {\bf 1995}, \emph{349},~11. 

\bibitem{kpe95} Knorren, R.; Prakash, M.; Ellis, P.J. Strangeness in hadronic stellar matter. 
\emph{Phys.~Rev.~C} {\bf 1995}, \emph{52},~3470. 

\bibitem{sm96} Schaffner, J.; Mishustin, I.N. Hyperon-rich matter in neutron stars. 
\emph{Phys.~Rev.~C} {\bf 1996}, \emph{53},~1416. 

\bibitem{pal2000} Pal, S.; Bandyopadhyay, D.; Greiner, W. Antikaon condensation in neutron stars.  \emph{Nucl.~Phys.~A} {\bf 2000}, \emph{674},~553.

\bibitem{m2008} Muto, T. Interplay between kaon condensation and hyperons in highly dense matter. \emph{Phys.~Rev.~C} {\bf  2008}, \emph{77},~015810.

\bibitem{mishra2010} Mishra, A.; Kumar, A.; Sanyal, S.; Dexheimer, V.; Schramm, S. Kaon properties in (proto-)neutron star matter. \emph{Eur.~Phys.~J. A}~{\bf 2010}, \emph{45},~169.

\bibitem{cb2014} Char, P.; Banik, S. Massive neutron stars with antikaon condensates in a density-dependent hadron field theory. \emph{Phys.~Rev.~C} {\bf 2014}, \emph{90},~015801. 

\bibitem{mbb2021} Malik, T.; Banik, S.; Bandyopadhyay, D. Equation-of-state Table with Hyperon and Antikaon for Supernova and Neutron Star Merger. \emph{Astrophys.~J.}~{\bf 2021}, \emph{910}, 96.

\bibitem{ma2023} Ma, F.; Wu, C.; Guo, W. Kaon-meson condensation and $\Delta$ resonance in hyperonic stellar matter within a relativistic mean-field model. \emph{Phys.~Rev.~C}  {\bf 2023}, \emph{107},~045804.

\bibitem{mmt2021} Muto, T.; Maruyama, T.; Tatsumi, T. Effects of three-baryon forces on kaon condensation in hyperon-mixed matter. \emph{Phys.~Lett.~B} {\bf 2021}, \emph{820},~136587.

\bibitem{mmt2022} Muto, T.; Maruyama, T.; Tatsumi, T. Kaon-baryon coupling schemes and kaon condensation in hyperon-mixed matter. 
\emph{Prog.~Theor.~Exp.~Phys.}~{\bf 2022}, \emph{2022}~093D03. 

\bibitem{muto2024} Muto, T. Properties of kaon-condensed phase in hyperon-mixed matter with three-baryon forces.arXiv:{\bf 2024} 2411.09967v1.

\bibitem{muto2002} {Muto, T. Kaonic modes in hyperonic matter and p-wave kaon condensation. \emph{Nucl.~Phys.~A} {\bf  2002}, \emph{697},~225. } 

\bibitem{PDG2020} Zyla, P.A.; Barnett, R.M.; Beringer, J.; Dahl, O.; Dwyer,D.A.; Groom,D.E.; Lin, C. -J.;  Lugovsky,K.S.; Pianori, E.; Robinson,D.J. Review of Particle Physics. \emph{Prog.~Theor.~Exp.~Phys.}~{\bf 2020}, \emph{2020},~083C01.

\bibitem{cohen1992} Cohen, T.D.; Furnstahl, R.J.; Griegel, D.K. Quark and gluon condensates in nuclear matter. \emph{Phys.~Rev.~C} {\bf 1992},  \emph{45},~1881.

\bibitem{t2008} Tamagaki, R. Universal short-range repulsion in the baryon system originating from the confinement: Approach in string-junction model. \emph{Prog.~Theor.~Phys.}~{\bf 2008}, \emph{119}, 965.

\bibitem{tnt2008} Takatsuka, T.; Nishizaki, S.; Tamagaki, R. Universal three???body repulsion suggested by neutron stars. \emph{AIP~Conf.~Proc.}~{\bf 2008}, \emph{1011}, 209.

\bibitem{nth1994} Nishizaki, S.; Takatsuka, T.; Hiura, J. Properties of hot asymmetric nuclear matter.\emph{Prog.~Theor.~Phys.}~{\bf 1994}, \emph{92}, 93.

\bibitem{lp1981}Lagaris, I.E.; Pandharipande, V.R. Variational calculations of realistic models of nuclear matter. \emph{Nucl.~Phys.~A} {\bf 1981}, \emph{359}, 349.

\bibitem{oertel2017} Oertel, M.; Hempel, M.; K\"ahn, T.; Typel, S. Equations of state for supernovae and compact stars. \emph{Rev.~Mod.~Phys.}~{\bf 2017}, \emph{89},~015007. 

\bibitem{sdg94} Schaffner, J.; Dover, C.B.; Gal, A.; Greiner, C.; Millener, D.J.; St\"ocker, H. Multiply strange nuclear systems. \emph{Ann.~Phys.}~{\bf 1994}, \emph{235},~35.

\bibitem{ohki08} Ohki, H.; Fukaya, H.; Hashimoto, S.; Kaneko, T.; Matsufuru, H.;  Noaki, J.; Onogi, T.; Shintani, E.; Yamada, N,; et al. Nucleon sigma term and strange quark content from lattice QCD with exact chiral symmetry. \emph{Phys.~Rev.~D} {\bf 2008}, \emph{78},~054502.

\bibitem{Durr2016} Durr, S.; Fodor, Z.; Hoelbling, C.; Katz,S.D.; Krieg,S.; Lellouch,L.; Lippert,T.; Metivet,T.; Portelli, A.; et al. Lattice computation of the nucleon scalar quark contents at the physical point. 
\emph{Phys.~Rev.~Lett.}~{\bf 2016}, \emph{116},~172001.

\bibitem{Alex2020} Alexandrou, C.; Bacchio, S.; Constantinou, M.; Finkenrath,J.; Hadjiyiannakou, K.; Jansen,K.; 
Koutsou,G.;Vaquero Aviles-Casco, A. Nucleon axial, tensor, and scalar charges and $\sigma$-terms in lattice QCD. \emph{Phys.~Rev. D}~{\bf 2020}, \emph{102},~054517.

\bibitem{gls1991} Gasser, J.; Leutwyler, H.; Sainio, M.E. Sigma-term update. \emph{Phys.~Lett.~B} {\bf 1991}, \emph{253},~260.

\bibitem{a2021} Alarc{\' o}n,  J.M. Brief history of the pion???nucleon sigma term.
\emph{Eur.~Phys.~J.~Spec.~Top.}~{\bf 2021}, \emph{230},~1609.

\bibitem{jk1987} Jaffe, R.L.; Korpa, C.L. The pattern of chiral symmetry breaking and the strange quark content of the proton. \emph{Comm.~Nucl.~Part.~Phys.}~{\bf 1987}, \emph{17},~163.

\bibitem{hk1991} Hatsuda, T.; Kunihiro, T. Flavor mixing in the low energy hadron dynamics: interplay of the SU$_f$(3) breaking and the U$_A$(1) anomaly. \emph{Z.~Phys.~C} {\bf 1991}, \emph{51},~49.

\bibitem{ro2000} Ramos, A.; Oset, E. The properties of $\bar K$ in the nuclear medium. \emph{Nucl.~Phys.~A} {\bf 2000}, \emph{671},~481.

\bibitem{wrw1997} Waas, T.; Rho, M.; Weise, W. Effective kaon mass in dense baryonic matter: role of correlations. \emph{Nucl.~Phys.~A} {\bf 1997}, \emph{617},~449.

\bibitem{ww1997} Waas, T.; Weise, W. $S$-wave interactions of $\bar K$ and $\eta$ mesons in nuclear matter. \emph{Nucl.~Phys.~A} {\bf 1997}, \emph{625},~287.

\bibitem{ichikawa2020} Ichikawa, Y.; Yamagata-Sekihara, J.; Ahn, J.K.; Akazawa, Y.; Aoki, K.; Botta, E.; Ekawa, H.; Evtoukhovitch, P.; Feliciello, A.; Fujita, M.; et al. An event excess observed in the deeply bound region of the $^{12}$C ($K^-$, $p$) missing-mass spectrum. \emph{Prog.~Theor.~Exp.~Phys.} {\bf 2020}, \emph{2020}, 123D01.

\bibitem{sty2013} Shanahan, P.E.; Thomas, A.W.; Young, R.D. Sigma terms from an SU(3) chiral extrapolation. \emph{Phys.~Rev.~D} {\bf 2013}, \emph{87},~074503.

\bibitem{Lutz2014} Lutz, M.F.M.; Bavontaweepanya, R.; Kobadaj, C.; Schwarz, K. Finite volume effects in the chiral extrapolation of baryon masses. \emph{Phys.~Rev.~D} {\bf 2014}, \emph{90},~054505.

\bibitem{mti2000} Muto, T.; Tatsumi, T.; Iwamoto, N. Nonequilibrium weak processes in kaon condensation. I. Reaction rate for the thermal kaon process. \emph{Phys.~Rev.~D} {\bf 2000}, \emph{61},~063001.

\bibitem{mti2000-2} Muto, T.; Tatsumi, T.; Iwamoto, N. Nonequilibrium weak processes in kaon condensation. II. Kinetics of condensation. \emph{Phys.~Rev.~D} {\bf 2000}, \emph{61},~083002.

\bibitem{BPS1971} Baym, G.; Pethick, C.; Sutherland, P. The ground state of matter at high densities: Equation of state and stellar models. \emph{Astrophys.~J.}~{\bf 1971}, \emph{170},~299.

\bibitem{demo10} Demorest, P.B.; Pennucci, T.; Ransom, S.M.; Roberts, M.S.E.; Hessels, J.W.T. A two-solar-mass neutron star measured using Shapiro delay. \emph{Nature}~{\bf 2010}, \emph{467},~1081.

\bibitem{fonseca2016} Fonseca, E.; Pennucci, T.T.; Ellis, J.A.; Stairs, I.H.; Nice, D.J.; Ransom, S.M.; Demorest, P.B.; Arzoumanian, Z.; Crowter, K.; Dolch, T.; et al.The nanograv nine-year data set: mass and geometric measurements of binary millisecond pulsars. \emph{Astrophys.~J.}~{\bf 2016}, \emph{832},~167.

\bibitem{ant13} Antoniadis, J.; Fereire, P.C.C.; Wex, N.; Tauris, T.M.; Lynch, R.S.; van Kerkwijk, M.H.;Kramer, M.; Bassa, C.; Dhillon, V.S.;Driebe, T.;et al. A massive pulsar in a compact relativistic binary. \emph{Science}~{\bf 2013}, \emph{340},~448.

\bibitem{c2020} Cromartie, H.T.; Fonseca, E.; Ransom, S.M.; Demorest, P.B.; Arzoumanian, Z.; Blumer,H.; Brook, P. R.; DeCesar, M.E.; Dolch,T.; Ellis,J.A.;et al. Relativistic Shapiro delay measurements of an extremely massive millisecond pulsar. \emph{Nat.~Astron.}~{\bf 2020}, \emph{4},~72.

\bibitem{fonseca2021} Fonseca, E.; Cromartie, H.T.; Pennucci, T.T.; Ray, P.S.;  Kirichenko, A.Y.; Ransom, S.M.; Demorest, P.B.; Stairs, I.H.; Arzoumanian, Z.;Guillemot, L.; et~al. Refined mass and geometric measurements of the high-mass PSR J0740+6620. \emph{Astrophys.~J.~L.}~{\bf 2021}, \emph{915},~L12. \\ arXiv: {\bf 2021} 2104.00880.

\bibitem{miller2021} Miller, M.C.; Lamb, F.K.; Dittmann, A.J.; Bogdanov, S.; Arzoumanian, Z.; Gendreau, K.C.; 
Guillot, S.; Ho, W.C.G.; Lattimer, J.M.; Loewenstein, M.; et al. The radius of PSR J0740+6620 from NICER and XMM-Newton data. \emph{Astrophys.~J.~L.}~{\bf 2021}, \emph{918},~L28. arXiv:{\bf 2021} 2105.06979.

\bibitem{riley2021} Riley, T.E.; Watts, A.L.; Ray, P.S.; Bogdanov, S.; Guillot, S.; Morsink, S.M.; Bilous, A.V.; 
Arzoumanian,Z.; Choudhury, D.;  Deneva, J.S.; et al. A NICER view of the massive pulsar PSR J0740+6620 informed by radio timing and XMM-Newton spectroscopy. \emph{Astrophys.~J.~L.}~{\bf 2021}, \emph{918}~L27.~arXiv:{\bf 2021} 2105.06980.

\bibitem{riley2019} Riley, T.E.; Watts, A.L.; Bogdanov, S.; Ray, P.S.;  Ludlam, R.M.; Guillot,S.;  Arzoumanian, Z.; 
Baker, C.L.;  Bilous, A.V.; Chakrabarty, D.; et al. A NICER view of PSR J0030+0451: millisecond pulsar parameter estimation. \emph{Astrophys.~J.~L.}~{\bf 2019}, \emph{887},~L21.

\bibitem{miller2019} Miller, M.C.;  Lamb,F.K.; Dittmann, A.J.; Bogdanov, S.; Arzoumanian, Z.;  Gendreau, K.C.; Guillot,S.; Harding,A.K.;  Ho, W. C. G.; Lattimer, J. M.; et al. PSR J0030+0451 mass and radius from NICER data and implications for the properties of neutron star matter. \emph{Astrophys.~J.~L.}~{\bf 2019}, \emph{887},~L24.


\bibitem{dll2002} Danielewicz, P.; Lacey, R.; Lynch, W.G. Determination of the equation of state of dense matter. \emph{Science}~{\bf 2002}, \emph{298},~1592.

\bibitem{mannarelli2019} Mannarelli, M. Meson Condensation. \emph{Particles}~{\bf 2019}, \emph{2019},~411.

\bibitem{a2009} Abuki, H.; Anglani, R.; Gatto, R.; Pellicoro, M.; Ruggieri, M. Fate of pion condensation in quark matter: From the chiral limit to the physical pion mass. \emph{Phys.~Rev.~D} {\bf 2009}, \emph{79},~034032. 

\bibitem{kkz2019} Khunjua, T.G.; Klimenko, K.G.; Zhokhov, R.N. Charged pion condensation in dense quark matter: Nambu--Jona-Lasinio model study. \emph{Symmetry}~{\bf 2019},~\emph{2019},~778.

\bibitem{kkz2020} Khunjua, T.G.; Klimenko, K.G.; Zhokhov, R.N. Electrical neutrality and $\beta$-equilibrium conditions in dense quark matter: generation of charged pion condensation by chiral imbalance. 
\emph{Eur.~Phys.~J.~C} {\bf 2020}, \emph{80}, 995.

\bibitem{aa2020} Adhikari, P.; Andersen, J.O. Quark and pion condensates at finite isospin density in chiral perturbation theory. \emph{Eur.~Phys.~J.~C} {\bf 2020}, \emph{80}, 1028.

\bibitem{aa2020-2} Adhikari, P.; Andersen, J.O. Pion and kaon condensation at zero temperature in three-flavor 
$\chi$PPT at nonzero isospin and strange chemical potentials at next-to-leading order.
 \emph{JHEP}~{\bf 2020}, \emph{06},~170.

\bibitem{sst2011} Schmitt, A.; Stetina, S.; Tachibana, M. Ginzburg - Landau phase diagram for dense matter with axial anomaly, strange quark mass, and meson condensation. \emph{Phys.~Rev.~D} {\bf 2011}, \emph{83},~045008. 

\bibitem{mht2013} Masuda, K.; Hatsuda, T.; Takatsuka, T. Hadron-quark crossover and massive hybrid stars with strangeness. \emph{Astrophys.~J.}~{\bf 2013}, \emph{764},~12.

\bibitem{mht2016} Masuda, K.; Hatsuda, T.; Takatsuka, T. Hadron--quark crossover and hot neutron stars at birth. \emph{PTEP}~{\bf 2016}, \emph{2016},~021D01.

\bibitem{baym2019} Baym, G.; Furusawa, S.; Hatsuda, T.; Kojo, T.; Togashi, H. New neutron star equation of state with quark???hadron crossover. \emph{Astrophys.~J.}~{\bf 2019}, \emph{885},~42.

\bibitem{kojo2021} Kojo, T. Stiffening of matter in quark-hadron continuity. \emph{Phys.~Rev.~D} {\bf 2021}, \emph{104},~074005.

\bibitem{fujimoto2022} Fujimoto, Y.; Fukushima, K.; McLerran, L.D.; Praszalowicz, M. Trace anomaly as signature of conformality in neutron stars. 
\emph{Phys.~Rev.~Lett.}~{\bf 2022}, \emph{129},~252702.

\bibitem{tn2004} Tatsumi, T.; Nakano, E. arXiv:{\bf 2004} hep-ph/0408294v1. 

\bibitem{nt2005} Nakano, E.; Tatsumi, T. Chiral symmetry and density waves in quark matter. \emph{Phys.~Rev.~D} {\bf 2005}, \emph{71},~114006).

\bibitem{son2000} Son, D.T.; Stephanov, M. Inverse meson mass ordering in the color-flavor-locking phase of high-density QCD. \emph{Phys.~Rev.~D} {\bf 2000}, \emph{61},~07402.

\bibitem{bedaque2002} Bedaque, P.; Schafer, T. High-density quark matter under stress. \emph{Nucl.~Phys.~A} {\bf 2002}, \emph{697},~802.

\bibitem{kr2002} Kaplan, D.B.; Reddy, S. Novel phases and transitions in color flavor locked matter. \emph{Phys.~Rev.~D} {\bf 2002}, \emph{65},~054042.

\end{thebibliography}
\end{document}